\documentclass[11pt,a4paper]{article}

\usepackage{jheppub}
\usepackage{latexsym}
\usepackage{multirow}
\usepackage{color}
\usepackage[usenames,dvipsnames,table]{xcolor}
\usepackage{graphicx}
\usepackage{epsfig}  
\usepackage{epsf}    
\usepackage{dcolumn}
\usepackage{bm}
\usepackage{dcolumn}
\usepackage{textcomp}
\usepackage{float}
\usepackage{subfig}

\usepackage[toc,page]{appendix}

\usepackage{lineno}

\usepackage{hypcap}
\usepackage[]{hyperref}
\usepackage{makecell}
\usepackage{color}
\usepackage{pifont}
\usepackage{appendix}
\usepackage{amsmath}
\usepackage{multirow,bigdelim}  
\usepackage{lineno}
\usepackage[normalem]{ulem}
\hypersetup{
  bookmarks=true,         
  unicode=false,          
  pdftoolbar=true,        
 pdfmenubar=true,        
 pdffitwindow=true,     
 pdfstartview={FitH},    
 pdfsubject={Neutrino Oscillations Phenomenology},   
 pdfnewwindow=true,      
 pdfcreator={RevTeX},
 colorlinks=true,       
 linkcolor=red,          
 citecolor=blue,        
 filecolor=black,      
 urlcolor=blue,           
  }

\newcommand{\be}{\begin{equation}}
\newcommand{\ee}{\end{equation}}
\newcommand{\ba}{\begin{eqnarray}}
\newcommand{\ea}{\end{eqnarray}}

\newcommand{\capdef}{}
\newcommand{\mycaption}[2][\capdef]{\renewcommand{\capdef}{#2}
       \caption[#1]{{\footnotesize #2}}}
\makeatletter
\renewcommand{\fnum@table}{\textbf{\tablename~\thetable}}
\renewcommand{\fnum@figure}{\textbf{\figurename~\thefigure}}
\makeatother

\preprint{IP/BBSR/2022-07, TIFR/TH/22-40}

\title{Locating the Core-Mantle Boundary using Oscillations of Atmospheric Neutrinos}

\author[a,b]{Anuj Kumar Upadhyay,}
\author[b,c,d]{Anil Kumar,}
\author[b,d,e]{Sanjib Kumar Agarwalla,}
\author[f]{Amol Dighe} 

\affiliation[a]{Department of Physics, Aligarh Muslim University, Aligarh-202002, India }
\affiliation[b]{Institute of Physics, Sachivalaya Marg, Sainik School Post,
  Bhubaneswar 751005, India}
\affiliation[c]{Applied Nuclear Physics Division, Saha Institute of
  Nuclear Physics, Block AF, Sector 1, Bidhannagar, Kolkata 700064, India}
\affiliation[d]{Homi Bhabha National Institute, Anushakti Nagar,
  Mumbai 400094, India}
\affiliation[e]{Department of Physics \& Wisconsin IceCube Particle Astrophysics Center, University of Wisconsin, Madison, WI 53706, U.S.A}
\affiliation[f]{Tata Institute of Fundamental Research, Homi Bhabha Road,
	Colaba, Mumbai 400005, India}

\emailAdd{anuju@iopb.res.in  (ORCID: 0000-0003-1957-2626)}
\emailAdd{anil.k@iopb.res.in (ORCID: 0000-0002-8367-8401)}
\emailAdd{sanjib@iopb.res.in (ORCID: 0000-0002-9714-8866)}
\emailAdd{amol@theory.tifr.res.in (ORCID: 0000-0001-6639-0951)}

\abstract{
Atmospheric neutrinos provide a unique avenue to explore the internal structure of Earth based on weak interactions, which is complementary to seismic studies and gravitational measurements. In this work, we demonstrate that the atmospheric neutrino oscillations in the presence of Earth matter can serve as an important tool to locate the core-mantle boundary (CMB). An atmospheric neutrino detector like the proposed 50 kt magnetized ICAL at INO can observe the core-passing neutrinos efficiently. These neutrinos would have experienced the MSW resonance and the parametric or neutrino oscillation length resonance. The net effect of these resonances on neutrino flavor conversions depends upon the location of CMB and the density jump at that radius. We quantify the capability of ICAL to measure the location of CMB in the context of multiple three-layered models of Earth. For the model where the density and the radius of core are kept flexible while the mass and radius of Earth as well as the densities of outer and inner mantle are fixed, ICAL can determine the location of CMB with a 1$\sigma$ precision of about 250 km with an exposure of 1000 kt$\cdot$yr. With the 81-layered PREM profile, this $1\sigma$ precision would be about 350 km. The charge identification capability of ICAL plays an important role in achieving this precision.	  
}

\arxivnumber{2211.08688}

\begin{document}
\maketitle
\flushbottom

\section{Introduction and Motivation}
\label{sec:introduction}
The exact nature of deep interior of Earth has been a long-standing puzzle. The internal regions of Earth are inaccessible due to extreme temperatures and pressures. Direct measurements are not feasible after a depth of a few km. Most of the information available about the internal structure of Earth has been obtained using indirect probes such as the gravitational measurements~\cite{Ries:1992,Luzum:2011,Rosi:2014kva,astro_almanac,Williams:1994,Chen:2014} and the seismic studies~\cite{Gutenberg:1914,Robertson:1966,Volgyesi:1982,Loper:1995,Alfe:2007,Stacey_Davis:2008,McDonough:2022,Thorne:2022,Hirose:2022}. For example, gravitational measurements provide information on the mass ~\cite{Ries:1992,Williams:1994,Luzum:2011,Rosi:2014kva,astro_almanac}, and the moment of inertia of Earth~\cite{Williams:1994,Chen:2014}. These tell us that the average density of Earth deep below must be much larger than the typical rock density at the surface. 
 
Seismic waves are generated when the tectonics plates in the outermost layers of Earth slide against each other. The origin of earthquake, called epicenter, is typically located up to a depth of about 700 km~\cite{Stacey_Davis:2008,Volgyesi:1982}. These seismic waves travel in all directions through the bulk of Earth and can reach the surface all over the globe. They are measured by seismometers which provide the information about the time, location, and intensity of the earthquake.  While passing though Earth, these waves can get reflected or refracted on encountering matter with different densities and composition~\cite{Volgyesi:1982}. The features observed in the seismic waves are used to infer the internal structure of Earth.

From seismic studies~\cite{Gutenberg:1914,Robertson:1966,Volgyesi:1982,Loper:1995,Alfe:2007,Stacey_Davis:2008,McDonough:2022,Thorne:2022,Hirose:2022}, we know that the Earth consists of a layered structure in the form of concentric shells. As we go from outside to deep inside the Earth, the layers are arranged in the order of increasing densities as crust, outer mantle, inner mantle, outer core and inner core. The density changes sharply at the core-mantle boundary (CMB) which has been infered to be $3483 \pm 5$ km using seismological measurements~\cite{Gutenberg:1914,McDonough:2003}. Even though seismic studies and gravitational measurements have provided us a huge amount of data and revealed crucial information about the internal structure of Earth, there are still many open issues~\cite{McDonough:2022}. For example, the radius, mass and chemical composition of the core are not very well known. The density jump between outer core and inner core is not known precisely.  There are many unexplained structures and heterogeneities observed in the lowermost mantle, beneath Africa and Pacific, that show lower-than-average seismic wave speeds which are known as large low-shear-velocity provinces (LLSVPs) and ultra low velocity zones (ULVZs)~\cite{Garnero:2016,McNamara:2019,Thorne:2022}. Active research is being pursued to determine the composition of the bulk silicate Earth~\cite{McDonough:1995,McDonough:2022}. If we talk about the presence of light elements inside Earth, we don't know how much H\textsubscript{2}O is present in the mantle and how much H is present in the core~\cite{Williams:2001,Hirose:2021,Hirose:2022}. The precise measurement of radioactive power produced in the mantle and core is important to understand the thermal dynamics inside Earth~\cite{Araki:2005qa,Borexino:2015ucj,Bellini:2021sow,KamLAND:2022vbm}.

Complementary information using additional probes such as neutrinos can improve our understanding about the structure of Earth. Neutrinos interact only via weak interactions, which enables most of the neutrinos to pass through Earth without getting absorbed. The cross section for interaction of neutrino with nucleons increases with energy. At energies higher then a few TeV, the interaction cross-section is large enough to have sizable absorption of neutrinos inside the Earth~\cite{Gandhi:1995tf,IceCube:2017roe}.  The idea of using the attenuation in the flux of neutrinos to probe the internal structure of Earth was proposed in Ref.~\cite{Placci:1973,Volkova:1974xa}, and detailed studies using neutrinos from various sources, such as man-made neutrinos~\cite{Placci:1973,Volkova:1974xa,Nedyalkov:1981,Nedyalkov:1981yy,Nedialkov:1983,Krastev:1983,DeRujula:1983ya,Wilson:1983an,Askarian:1984xrv,Volkova:1985zc,Tsarev:1985yub,Borisov:1986sm,Tsarev:1986xg,Borisov:1989kh,Winter:2006vg}, extraterrestrial neutrinos~\cite{Wilson:1983an,Kuo:1995,Crawford:1995,Jain:1999kp,Reynoso:2004dt}, and atmospheric neutrinos~\cite{Gonzalez-Garcia:2007wfs,Borriello:2009ad,Takeuchi:2010,Romero:2011zzb} have been performed. The analysis in Ref.~\cite{Gonzalez-Garcia:2007wfs} made a forecast that using the absorption of high-energy atmospheric neutrinos (in the range of 10 to 100 TeV), IceCube can reject the homogeneity of the Earth at 3.4$\sigma$ confidence level in 10 years with conservative assumptions on the theoretical and systematic uncertainties. The analysis of the one-year data of multi-TeV atmospheric muon neutrinos at IceCube, carried out in Ref.~\cite{Donini:2018tsg}, estimated the densities of various Earth layers using the absorption of high-energy neutrinos for the first time. They determined the mass of the Earth to be $M^\nu_E = 6.0^{+1.6}_{-1.3} \times 10^{24}$ kg, which is in good agreement with the gravitational measurement. Another way of exploring the interior of Earth can be the use of diffraction patterns produced by the coherent scattering of neutrinos with the Earth's matter, but this is not feasible with the present technology~\cite{Fortes:2006}.

The improvement in the precision measurement of neutrino oscillation parameters~\cite{Capozzi:2021fjo,NuFIT,Esteban:2020cvm,deSalas:2020pgw}, including reactor mixing angle $\theta_{13}$~\cite{Meregaglia:2017spw,DayaBay:2018yms,RENO:2018dro,RENO:2022}, have provided a new way to explore Earth's interior via matter effects in neutrino oscillations in the multi-GeV range of energies. The matter effects come into picture during the interactions of upward-going atmospheric neutrinos with the ambient electrons present inside Earth. This charged-current (CC) coherent elastic forward scattering results into an effective matter potential for neutrinos which depends upon the number density of electrons along the neutrino trajectory. The possibility of probing the internal structure of Earth using these density-dependent matter effects is known as ``neutrino oscillation tomography''. These studies have been performed using man-made neutrino beams~\cite{Ermilova:1986ph,Nicolaidis:1987fe,Ermilova:1988pw,Nicolaidis:1990jm,Ohlsson:2001ck,Ohlsson:2001fy,Winter:2005we,Minakata:2006am,Gandhi:2006gu,Tang:2011wn,Arguelles:2012nw}, supernova neutrinos~\cite{Akhmedov:2005yt,Lindner:2002wm}, solar neutrinos~\cite{Ioannisian:2002yj,Ioannisian:2004jk,Akhmedov:2005yt,Ioannisian:2015qwa,Ioannisian:2017chl,Ioannisian:2017dkx,Bakhti:2020tcj}, and atmospheric neutrinos~\cite{Agarwalla:2012uj,IceCube-PINGU:2014okk,Rott:2015kwa,Winter:2015zwx,Bourret:2017tkw,Bourret:2019wme,Bourret:2020zwg,DOlivo:2020ssf,Kumar:2021faw,Kelly:2021jfs,Maderer:2021aeb,Capozzi:2021hkl,Denton:2021rgt,Upadhyay:2021kzf,Maderer:2022toi,DOlivoSaez:2022vdl}. In the sub-GeV and multi-GeV energy ranges, atmospheric neutrinos are the best source of neutrinos to probe the internal structure of Earth because they have access to a wide range of baselines starting from about 15 km to 12750 km which cover all the layers of Earth. Sensitivity studies for the current and future atmospheric neutrino experiments like IceCube~\cite{Rott:2015kwa}, Precision IceCube Next Generation Upgrade (PINGU)~\cite{IceCube-PINGU:2014okk}, Oscillation Research with Cosmics in the Abyss (ORCA)~\cite{KM3Net:2016zxf}, Deep Underground Neutrino Experiment (DUNE)~\cite{DUNE:2021tad}, Hyper-Kamiokande (Hyper-K)~\cite{Hyper-Kamiokande:2018ofw}, and Iron Calorimeter (ICAL) detector~\cite{Kumar:2021faw} have been performed. These studies have shown how to detect the presence of core-mantle boundary using ICAL~\cite{Kumar:2021faw}, constrain the average densities of the core, and the mantle using ORCA~\cite{Winter:2015zwx,Maderer:2021aeb,Capozzi:2021hkl,DOlivoSaez:2022vdl} and DUNE~\cite{Kelly:2021jfs}, determine the position of the core-mantle boundary using DUNE~\cite{Denton:2021rgt}, and explore the chemical composition of the Earth's core using PINGU~\cite{IceCube-PINGU:2014okk}, Hyper-K and IceCube~\cite{Rott:2015kwa},  as well as ORCA~\cite{Bourret:2017tkw,Bourret:2019wme,Bourret:2020zwg,Maderer:2021aeb,Maderer:2022toi,DOlivoSaez:2022vdl}.

Further information about the chemical composition of the Earth can also be obtained using the geoneutrinos which are produced during the decay of radioactive elements such as Uranium, Thorium and Potassium~\cite{Araki:2005qa,Borexino:2015ucj,McDonough:2020,Bellini:2021sow,KamLAND:2022vbm,McDonough:2022a}. Geoneutrinos may shed light on the radiogenic contribution to the heat budget of Earth. Since neutrinos use weak interactions to probe the internal structure of Earth, which is complementary to seismic studies based on electromagnetic interactions and gravitational measurements based on gravitational interactions, this would pave the way for ``multi-messenger tomography of Earth''.

The proposed 50 kt ICAL detector at the upcoming India-based Neutrino Observatory (INO)~\cite{ICAL:2015stm} would be able to detect atmospheric neutrinos and antineutrinos in the multi-GeV range of energies and over a wide range of baselines. Thanks to the presence of magnetic field of about 1.5 T, ICAL would be able to distinguish between neutrino (by observing $\mu^-$) and antineutrino (by observing $\mu^+$) events separately. Further, due to its good directional resolution, ICAL would be able to observe and identify the neutrinos passing through core and mantle separately. The ICAL detector would be sensitive to Mikheyev-Smirnov-Wolfenstein (MSW) resonance~\cite{Wolfenstein:1977ue, Mikheev:1986gs, Mikheev:1986wj}, which occurs around the energies of 6 to 10 GeV for mantle-passing neutrinos. It can also observe neutrino oscillation length resonance (NOLR)~\cite{Petcov:1998su,Chizhov:1998ug,Petcov:1998sg,Chizhov:1999az,Chizhov:1999he} or parametric resonance (PR)~\cite{Akhmedov:1998ui,Akhmedov:1998xq}, which occurs around the energies of 3 to 6 GeV for some of the core-passing neutrinos. The NOLR/ parametric resonance occurs due to a sudden jump in the density when we go from mantle to core. The pattern of NOLR/ parametric resonance depends upon the amount of density jump and the location of CMB. In the present work, we explore the impact of modifying the CMB radius on neutrino oscillations and hence, on the reconstructed muon events at the ICAL detector. We demonstrate that the location of core-mantle boundary can be measured using the matter effects in atmospheric neutrino oscillations with the help of the ICAL detector. 

In section~\ref{sec:earth_model}, we discuss the internal structure of Earth and the propagation of seismic waves through different regions of Earth. In section~\ref{sec:CMB_variation}, we describe our methodology for exploring various modified-CMB scenarios with a simple three-layered density profile of Earth. The impact of CMB modification on neutrino oscillation probabilities is discussed in section~\ref{sec:probability} and \ref{sec:oscillograms}. The neutrino event simulation at ICAL is described in section~\ref{sec:events}. The method of statistical analysis to quantify the impact of CMB modification is explained in section~\ref{sec:statistical analysis}. In section~\ref{sec:results}, we evaluate the sensitivity of ICAL for determining the location of CMB. Finally, we present the summary and concluding remarks of this study in section~\ref{sec:conclusion}. In appendix~\ref{app:Case-II-variation}, we demonstrate the possible origin of asymmetric and non-monotonic behavior of sensitivities for modified-CMB scenarios with respect to the standard CMB in three-layered profile at the probability level. In appendix~\ref{app:Case-II-prem81}, we present the sensitivities for determining the location of CMB using the 81-layered PREM profile.

\section{A Brief Review of the Internal Structure of Earth}
\label{sec:earth_model}

The surface of Earth consists of soil, sand, rocks, mountains, rivers, oceans, ice, and lava etc, which differ from each other geologically as well as in terms of chemical composition. The layers beneath these structures are mostly solid and consist of various types of rocks. The direct observations of layers inside Earth have been possible only up to a few kms because the deepest hole in the world till today, with current technology, is only about 12 km which was drilled on the rocks of Kola Peninsula in the Murmansk region of Russia~\cite{Kozlovsky:1984,Kola_Superdeep}. Below the depth of 6 km, a temperature gradient of 20$^\circ$ C per km was observed instead of the expected gradient of 16$^\circ$ C per km. Eventually, the drilling had to stop due to extreme temperature of about 180$^\circ$ C. This exploration clearly shows that if we want to know more about the structure of Earth at the deeper locations, then we need to use indirect methods, for example, gravitational measurements~\cite{Ries:1992,Luzum:2011,Rosi:2014kva,astro_almanac,Williams:1994,Chen:2014}, studies of earthquakes or seismic waves~\cite{Gutenberg:1914,Robertson:1966,Volgyesi:1982,Loper:1995,Alfe:2007,Stacey_Davis:2008,McDonough:2022,Thorne:2022,Hirose:2022}, neutrino tomography~\cite{MMTE:2022}, etc.

Seismic waves are generated when the tectonic plates inside Earth slide and the energy is released in the form of vibrations. They get modified while propagating through Earth. The velocity and timing measurements of seismic waves are used to unravel the internal structure of Earth~\cite{Gutenberg:1914,Robertson:1966,Loper:1995,Alfe:2007}. Seismic studies have revealed that the Earth consists of the layers in the form of concentric shells which can be broadly divided into core and mantle. The radius of core is about half the radius of Earth ($R_E$) which is about 6371 km. The contribution of core and mantle to the mass of Earth are about 32\% and 68\%, respectively.

The information about Earth is also obtained using gravitational measurements. The mass of Earth ($M_E$) is gravitationally measured to be about $(5.9722\,\pm\,0.0006) \times 10^{24}$ kg~\cite{Ries:1992,Luzum:2011,Rosi:2014kva,astro_almanac} whereas its moment of inertia is about $(8.01736\,\pm\, 0.00097) \times 10^{37}$ kg m\textsuperscript{2}~\cite{Williams:1994,Chen:2014}. Since the measured moment of inertia is less than the expected one ($2/5 M_E R_E^2 \sim 9.7 \times 10^{37}$ kg m\textsuperscript{2}) for a uniform density distribution inside Earth, the density should be more as we go deeper inside the Earth. In other words, the matter should be concentrated more towards the center of Earth. Using $M_E$ and $R_E$, one can obtain the average density of Earth to be 5.5 g/cm\textsuperscript{3}, which is larger than the density of rocks ($\sim$2.8 g/cm\textsuperscript{3}) present on the surface. This observation also points towards the presence of regions of higher densities deep inside the Earth.

Now, let us try to understand, how the interiors of Earth are probed using seismology. Seismic waves can be classified into two categories -- shear (S) waves and pressure (P) waves. The S-waves result into the vibration of rocks perpendicular to the direction of propagation whereas the P-waves force the particles to vibrate along the direction of propagation. The S-waves and P-waves modify differently while passing through the layers of Earth. The P-wave can travel through both solid as well as liquid layers but the velocity of the P-wave decreases while passing through liquid layers. As far as the S-waves are concerned, they cannot travel through the liquid layers. 

\begin{figure}
	\centering
	\includegraphics[width=\linewidth]{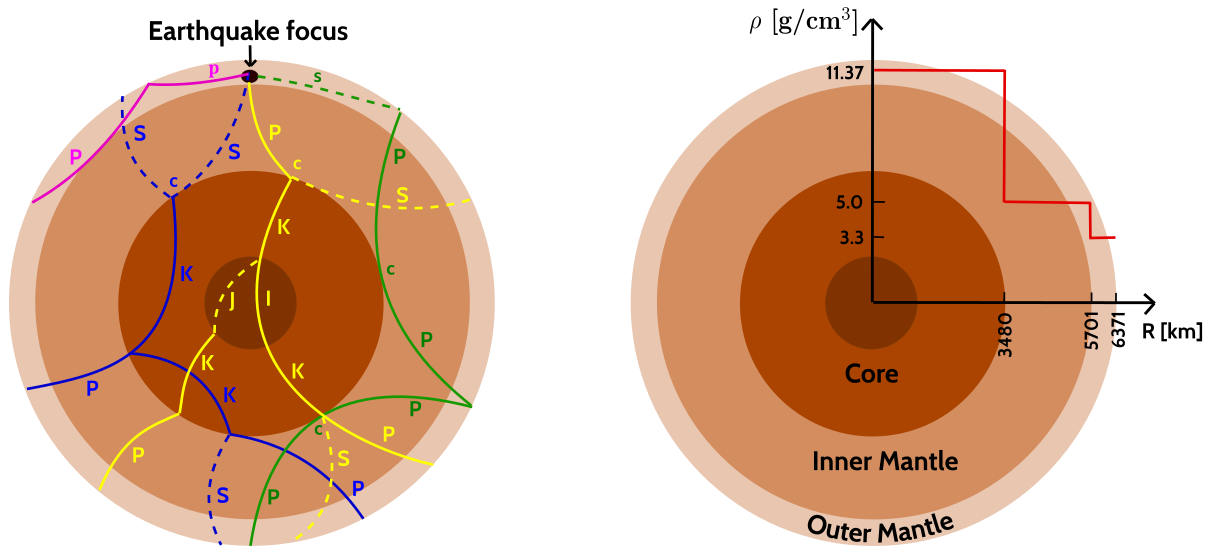}
	\mycaption{The left panel shows examples of seismic waves originating at the focus of an earthquake (black dot) and traveling inside Earth. The dashed and solid curves represent the S (shear) and P (pressure) waves, respectively. Colored curves represent the trajectories of these seismic waves. The components (S, P, K, I, J, s, p, c) of the seismic waves are described in the text. The right panel shows a three-layered model of Earth consisting of the core, inner mantle, and outer mantle. The red curve depicts the density profile of Earth as a function of radial distance from its center.}
	\label{fig:seismic_phases}
\end{figure}

The seismic waves originating from the center of the earthquake travel through the Earth and land at various positions on the surface. During their propagation through Earth, S and P-waves may get reflected and refracted at multiple density discontinuities inside the Earth, splitting into different segments as shown in the left panel of Fig.~\ref{fig:seismic_phases}. The nomenclature of these segments is as follows: 
\begin{itemize}
	\item S and P indicate the S-wave and P-wave traveling through the mantle.
	
	\item K and I stand for the P-waves passing through the outer and inner core, respectively.
	
	\item J indicates the segment of the S-wave traveling through the inner core.
	
	\item s and p indicate those S and P-waves, respectively, which initially travel upward from the center of the earthquake, and then get reflected downward from the Earth's outer surface.
	
	\item c represents an upward reflection at the core-mantle boundary.
\end{itemize}

The seismic studies indicate that the mantle consists of hot rocks of silicate~\cite{Robertson:1966}. The rocks in mantle are not molten. However, they are plastic in nature which allows them to change their shape over long timescales. Beneath the mantle, the Earth consists of a high density core which is composed of metals like iron and nickel. The inability of S-waves to pass through the outer core, and the decrease in the velocity of P-waves therein, indicate that the outer core is expected be liquid~\cite{Robertson:1966}. In 1936, I. Lehmann discovered the inner core by its higher P-wave velocity~\cite{Lehmann:1936}. Inner core is expected to be solid~\cite{Birch:1940}. The only possible reason why iron and other heavy metals may be solid at high temperatures inside the inner core would be because their melting points increase significantly at the tremendous pressure present there~\cite{Birch:1940}.

The data on the velocities of seismic waves are used to develop the Preliminary Reference Earth Model (PREM)~\cite{Dziewonski:1981xy} profile where the density of any layer is given as a one-dimensional function\footnote{The seismological studies indicate that the departure from spherical symmetry near CMB is small. For example, the ellipticity at CMB is about $2.5 \times 10^{-3}$ whereas the outer core surface topography is within 3 km~\cite{McDonough:2017}. Recently, several new three-dimensional Earth models like Shen-Ritzwoller  (S-R)~\cite{sr:2016}, FWEA18~\cite{FWEA18:2018}, SAW642AN~\cite{SAW642AN:2000}, CRUST1~\cite{CRUST1:2013}, etc. have been developed. At multi-GeV energies, the oscillation wavelength for atmospheric neutrinos is around a few thousands of km. Therefore, atmospheric neutrino oscillations are not expected to be sensitive to such detailed features.} 
of the radial distance  of the layer from the center of Earth. The PREM profile is mainly based on two empirical equations which relate the velocities of S and P-waves with the densities of the layers inside Earth. The first one is called the Birch's law~\cite{Birch:1964} and is valid for the outer mantle whereas the second one is known as Adams-Williamson equation~\cite{Williamson:1923} that is suitable for the inner mantle and the core. The parameters of these empirical relations depend upon temperature, pressure, composition, and elastic properties of the layers of Earth, which give rise to uncertainties in density distribution. The density of the mantle has a uncertainty of about 5\% whereas for the core, it is significantly larger~\cite{Bolt:1991,kennett:1998,Masters:2003}. 

The sharp change in the densities of layers results in the partial reflection and refraction of seismic waves. Such a sudden rise in density is also observed at the CMB. In the present work, we explore whether the matter effects in atmospheric neutrino oscillations can be used to probe the location of CMB. We consider a simple three-layered model of Earth which consists of core, inner mantle, and outer mantle as shown in the right panel of Fig.~\ref{fig:seismic_phases}. The core is present up to a radial distance of 3480 km from the center of Earth, the inner mantle spans from  3480 km to 5701 km, and the outer mantle is from 5701 km to the radius of Earth. The densities of core, inner mantle and outer mantle are taken as 11.37, 5, and 3.3 g/cm\textsuperscript{3}. Note that we have not considered the crust which has much smaller thickness as compared to the other layers.

\section{Location of CMB and Neutrino Oscillations}
\label{sec:CMB_variation_effect}

\subsection{Three-layered Models with Modified CMB Locations}
\label{sec:CMB_variation}

\begin{figure}
	\centering
	\includegraphics[width=1.0\linewidth]{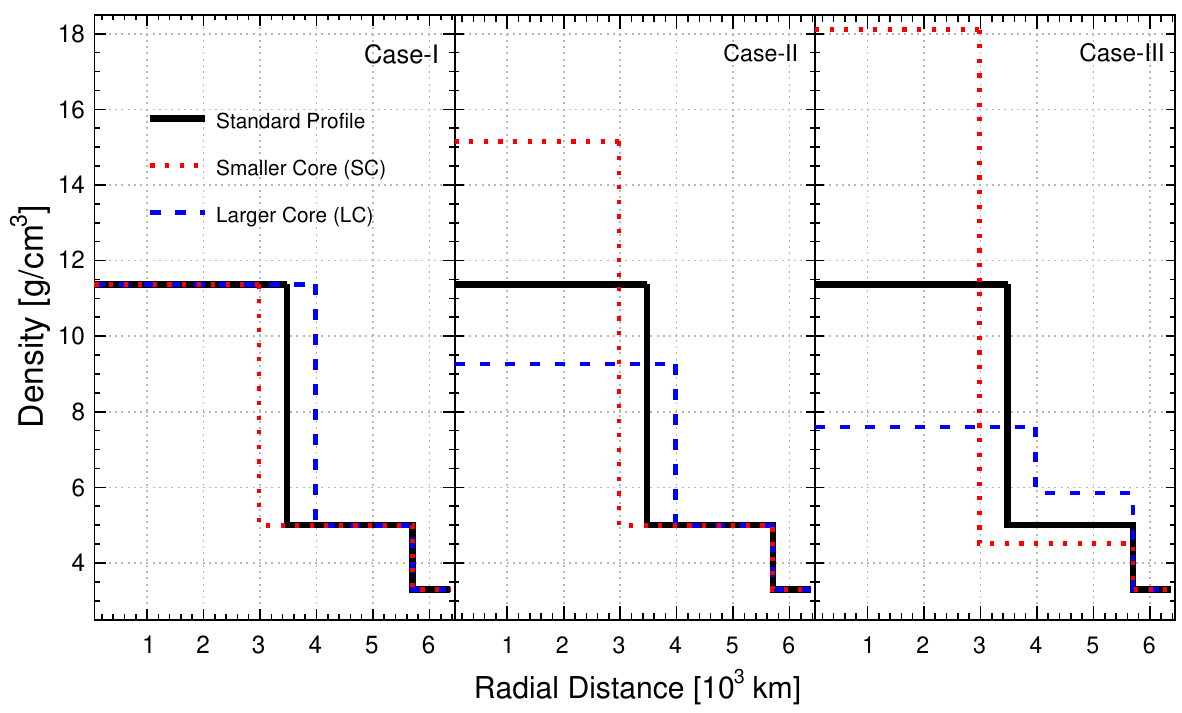}
	\mycaption{Densities as functions of radial distance for different ways of incorporating modified CMB in the three-layered profile of Earth. The black curves show the standard density profiles with $R_\text{CMB} = 3480$ km. The dotted-red (dashed-blue) curves indicate the density profile in the SC (LC) scenario, where $R_\text{CMB}$ in decreased (increased) by 500 km.  See text (Sec.~\ref{sec:CMB_variation}) for details of cases I, II, and III.
	 }
	\label{fig:CMB_variation_ways}
\end{figure}

\begin{table}
	\centering
	\begin{tabular}{| c | c |c c c |} 
		\hline
		\multirow{2}{*}{Three-layered profile} & Radius of  & \multicolumn{3}{c|}{ Layer densities (g/$\text{cm}^3$) } \\
		\cline{3-5}
		& layer boundaries (km) & core & inner mantle & outer mantle\\
		\hline\hline
		\textbf{Standard} & (3480, 5701, 6371) & 11.37 & 5.0 & 3.3 \\
		\hline
		\textbf{Case-I} &  & & &  \\
		Smaller core (SC) & (2980, 5701, 6371) & 11.37 & 5.0 & 3.3 \\ 
		Larger core (LC)  & (3980, 5701, 6371) & 11.37 & 5.0 & 3.3 \\ 
		\hline
		\textbf{Case-II} &  & & &   \\
		Smaller core (SC) & (2980, 5701, 6371) & 15.15 & 5.0 & 3.3 \\ 
		Larger core (LC) & (3980, 5701, 6371) & 9.26 & 5.0 & 3.3 \\ 
		\hline
		\textbf{Case-III} &  & & &  \\
		Smaller core (SC) & (2980, 5701, 6371) & 18.11 & 4.51 & 3.3\\ 
		Larger core (LC) & (3980, 5701, 6371) & 7.60 & 5.85 & 3.3 \\  
		\hline
	\end{tabular}
	\mycaption{ Boundaries of three layers and their densities in the three-layered profile of Earth when $R_\text{CMB}$ is modified by $\Delta R_\text{CMB}$. Smaller core (SC) stands for $\Delta R_\text{CMB} = - 500$ km, while larger core (LC) represents $\Delta R_\text{CMB} = + 500$ km. The corresponding modifications of density profiles are given by the cases I, II, III (see Sec.~\ref{sec:CMB_variation}).}
	\label{tab:CMB_variation}
\end{table}

In the present work, we use atmospheric neutrinos to probe the location of CMB where we explore how the matter effects in neutrino oscillations change in different modified-CMB scenarios. The standard CMB radius is taken to be $R_\text{CMB}$(standard)$=3480$ km. For illustration, we show the modification of the CMB radius by $\Delta R_\text{CMB} = -500$ km (smaller core or SC) and by $\Delta R_\text{CMB} = +500$ km (larger core or LC) in Fig~\ref{fig:CMB_variation_ways}.
In the standard, SC, and LC scenarios, the core spans the zenith angle range $\cos\theta_\nu \le -0.84$, $\cos\theta_\nu \le -0.88$, and $\cos\theta_\nu \le -0.78$, respectively.
 We further consider three different ways of modifying the density profile of the Earth. We term these as three ``cases'':
\begin{itemize}
	\item \textbf{Case-I:} The densities of all layers of  Earth are kept fixed. Note that the mass of Earth is not constrained in this case.
	
	\item \textbf{Case-II:} The densities of inner and outer mantle are kept fixed, while the density of core is modified to keep the mass of Earth invariant. 
	
	\item \textbf{Case-III:} The density of outer mantle is kept fixed, and the densities of core and inner mantle are modified, while keeping their individual masses invariant. It automatically ensures that the mass of Earth is unchanged.
\end{itemize}
Note that the Case-II is the most realistic one, since the density of the mantle is quite well measured by the combination of gravitational and seismological studies. However, we also analyze two dummy cases, Case-I and Case-III. In Case-I, we remove the constraint from the mass of Earth while imposing the constraint on the density of the core, compared to Case-II. In Case-III, we add the constraint on the mass of the mantle but remove the constraint on its density, compared to Case-II. This would allow us to obtain insights on the role of these constraints. In this study, we have not taken into account the constraint from the moment of inertia of Earth.

Table~\ref{tab:CMB_variation} presents the modified boundaries between various layers and their densities in the SC and LC scenarios. The densities of core and mantle may change depending upon the cases as explained above. We discuss the effect of $R_\text{CMB}$ modification on neutrino oscillations in the next section.

\subsection{Effect of Modified CMB Radius on Oscillation Probabilities}
\label{sec:probability}

\begin{table}
	\centering
	\begin{tabular}{|c|c|c|c|c|c|c|}
		\hline
		$\sin^2 2\theta_{12}$ & $\sin^2\theta_{23}$ & $\sin^2 2\theta_{13}$ & $\Delta m^2_
		\text{eff}$ (eV$^2$) & $\Delta m^2_{21}$ (eV$^2$) & $\delta_{\rm CP}$ & Mass Ordering\\
		\hline
		0.855 & 0.5 & 0.0875 & $2.49\times 10^{-3}$ & $7.4\times10^{-5}$ & 0 & Normal (NO)\\
		\hline 
	\end{tabular}
	\mycaption{The benchmark values of oscillation parameters considered in this analysis. These values are in good agreement with the present neutrino global fits~\cite{Capozzi:2021fjo,NuFIT,Esteban:2020cvm,deSalas:2020pgw}.}
	\label{tab:osc-param-value}
\end{table}

Atmospheric neutrinos and antineutrinos can be observed at ICAL separately in the mulit-GeV range of energies over a wide range of baselines from about 15 km to 12757 km. The upward-going neutrinos pass through the Earth and undergo charged-current interactions with the ambient electrons. This coherent forward scattering results in a matter potential $V_\text{CC}$ given by
\begin{equation}
V_\text{CC} = \pm\, \sqrt{2} G_F N_e \approx \pm \, 7.6 \times Y_e \times 10^{-14} \left[\frac{\rho}{\text{g/cm}^3}\right]~\text{eV}\,,
\end{equation}
where $G_F$ is Fermi coupling constant and $N_e$ is the ambient electron number density. Further, $Y_e = N_e/(N_p + N_n)$ is the relative number density of electron inside the matter having density $\rho$ where $N_p$ and $N_n$ denote the number densities of protons and neutrons. In this analysis, we assume that Earth is neutral and isoscalar, which implies $N_n \approx N_p = N_e$ and $Y_e = 0.5$. The positive and negative signs correspond to neutrinos and antineutrinos, respectively. Due to these opposite signs, the matter effects modify the oscillation patterns for neutrinos and antineutrinos differently. The charge identification capability of ICAL plays a crucial role in observing these different matter effects in neutrinos and antineutrinos separately, which enhances the sensitivity of ICAL towards locating the CMB as we demonstrate in our results. In the present work, we use the benchmark values of oscillation parameters given in Table~\ref{tab:osc-param-value}. The normal mass ordering (NO) represents $m_1 < m_2 < m_3$ whereas for inverted mass ordering (IO), $ m_3 < m_1 < m_2$. We implement NO vs. IO by taking opposite signs of the effective atmospheric mass-squared difference\footnote{The effective atmospheric mass-squared difference can be given in terms of $\Delta m^2_{31}$ and $\Delta m^2_{21}$ as follows~\cite{deGouvea:2005hk,Nunokawa:2005nx}
\begin{equation}
\Delta m^2_\text{eff} = \Delta m^2_{31} - \Delta m^2_{21} (\cos^2\theta_{12} - \cos \delta_\text{CP} \sin\theta_{13}\sin2\theta_{12}\tan\theta_{23}).
\label{eq:m_eff}
\end{equation}} $\Delta m^2_{\text{eff}}$.  

The matter effects result in a resonant enhancement in the effective value of the smallest lepton mixing angle $\theta_{13}$, which can become as large as $45^\circ$. This adiabatic resonance due to $\theta_{13}$-driven matter effects is known as the Mikheyev-Smirnov-Wolfenstein (MSW) resonance~\cite{Wolfenstein:1977ue,Mikheev:1986gs,Mikheev:1986wj}. For NO (IO), the MSW resonance occurs in neutrinos (antineutrinos).  For a given matter density $\rho$, the energy at which the MSW resonance occurs can be given as
\begin{align}
E_\text{res} & = \frac{\Delta m^2_{31} \cos 2\theta_{13}}{2\sqrt{2}G_F N_e} \simeq  \frac{\Delta m^2_{31}\left[\text{eV}^2\right] \cos 2\theta_{13}}{7.6 \times 10^{-14} \rho\left[\text{g/cm}^3\right]} ~ \text{eV}\,.
\end{align}
Now, if we consider the average density of mantle to be around 4.5 $\text{g/cm}^3$, then the MSW resonance occurs at approximately 7 GeV:
\begin{align}
E_\text{res} \simeq 7~\text{GeV} \left(\frac{4.5\,\text{g/cm}^3}{\rho}\right)\left(\frac{\Delta m^2_{31}}{2.4\times10^{-3}~\text{eV}^2}\right) \cos 2\theta_{13}\,. 
\end{align}
Therefore, the mantle-passing neutrinos feel the MSW resonance in the energy range of about 5 to 10 GeV. On the other hand, if we consider the density of core to be about 11.3 $\text{g/cm}^3$, the MSW resonance in the core occurs at approximately 2.8 GeV:
\begin{align}
E_\text{res} \simeq 2.8~\text{GeV} \left(\frac{11.3\,\text{g/cm}^3}{\rho}\right)\left(\frac{\Delta m^2_{31}}{2.4\times10^{-3}~\text{eV}^2}\right) \cos 2\theta_{13}\,. 
\end{align}
As we go along the trajectory of a core-passing neutrino, first the density increases from mantle to core and then it decrease from core to mantle. The quasi-periodic nature of densities along the neutrino path, combined with a sharp contrast in the densities of core and mantle at CMB,  may result in specific phase relationships between neutrino oscillation amplitudes in the core and the mantle for particular path lengths. This phenomenon is known as the neutrino oscillation length resonance (NOLR)~\cite{Petcov:1998su,Chizhov:1998ug,Petcov:1998sg,Chizhov:1999az,Chizhov:1999he} or parametric resonance (PR)~\cite{Akhmedov:1998ui,Akhmedov:1998xq}, which may enhance neutrino flavor conversions. 

\begin{figure}
	\centering
	
	\includegraphics[width=1.0\linewidth]{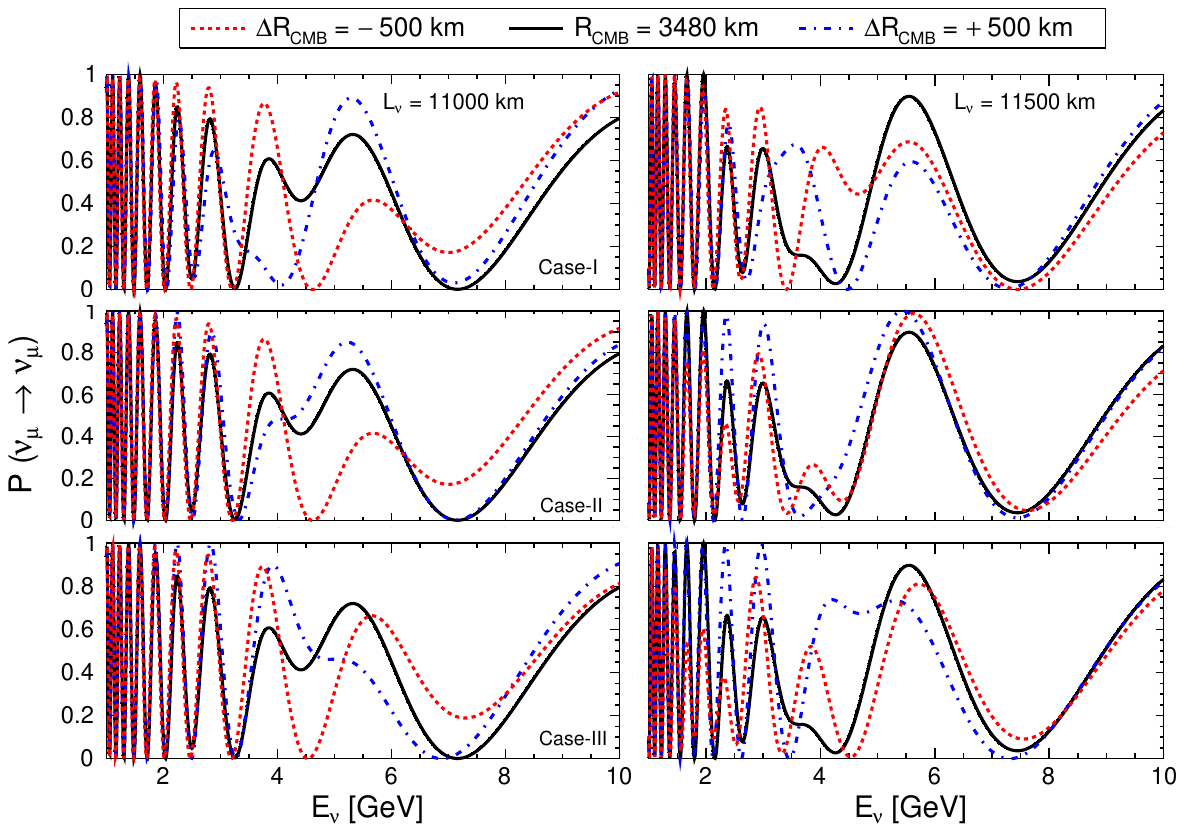}
	\mycaption{The three-flavor $\nu_\mu\rightarrow\nu_\mu$ survival probability as a function of neutrino energy $E_\nu$. The solid-black, dotted-red, and dashed-blue curves represent the scenarios of standard core ($R_\text{CMB} = 3480$ km), SC ($\Delta R_\text{CMB} = -\,500$ km), and LC ($\Delta R_\text{CMB} = +\,500$ km), respectively. The top, middle, and bottom panels correspond to the Case-I, Case-II, and Case-III, respectively. The left (right) panels consider the neutrino baseline $L_\nu$ of 11000 (11500) km. We use three-flavor neutrino oscillation parameters given in Table~\ref{tab:osc-param-value}, where we assume NO and $\sin^2\theta_{23}$ = 0.5.}
	\label{fig:Prob_case_1}
\end{figure}

The muon neutrino events at ICAL are contributed by both $\nu_\mu \rightarrow \nu_\mu$ survival as well as $\nu_e \rightarrow \nu_\mu$ appearance channels. However, since the contribution of the survival channel is more than 98\%, here we present the effect of $R_\text{CMB}$ modification only on the oscillation probabilities for $\nu_\mu \rightarrow \nu_\mu$ survival channel. Figure~\ref{fig:Prob_case_1} shows the three-flavor $\nu_{\mu}$ survival probability $P({\nu_\mu \rightarrow \nu_\mu})$ as a function of energy for NO. The top, middle, and bottom panels are for Case-I, Case-II, and Case-III, respectively. We consider the baseline of 11000 km (11500 km) in the left (right) panels. For the baseline of 11500 km, neutrinos pass through the core for all the three scenarios --- standard core ($R_\text{CMB} = 3480$ km), SC ($\Delta R_\text{CMB} = -\,500$ km), and LC ($\Delta R_\text{CMB} = +\,500$ km). However, for the baseline of 11000 km, neutrinos pass through the core in the scenarios where we have standard core and the larger core (LC). But in the scenario where we consider smaller core (SC), neutrinos do not pass through the core for the 11000 km baseline. This feature gives rise to additional modifications in neutrino oscillation probabilities for baselines around 11000 km and thus, contributes to the sensitivity for locating CMB. The $\theta_{13}$-driven matter resonances inside Earth at atmospheric frequency result in significant modification of $\nu_{\mu}$ survival probability, $P({\nu_\mu \rightarrow \nu_\mu})$.  The curves corresponding to SC and LC scenarios have different patterns with respect to each other for a given case and baseline. The oscillation patterns are also quite different in the three cases. The deviations in oscillation patterns mostly occur in the energy range of 3 to 6 GeV, which corresponds to the NOLR/parametric resonance. These observations imply that the neutrinos are highly sensitive to the modification of CMB as well as the way of modifying the density profile. Further, a comparison of the left and right panels shows that the modifications in neutrino oscillation patterns due to $R_\text{CMB}$ modification also depend upon the baselines through which neutrinos have passed.

\subsection{Effect of Modified CMB Radius on Oscillograms}
\label{sec:oscillograms}

\begin{figure}
	\centering
	\includegraphics[width=\linewidth]{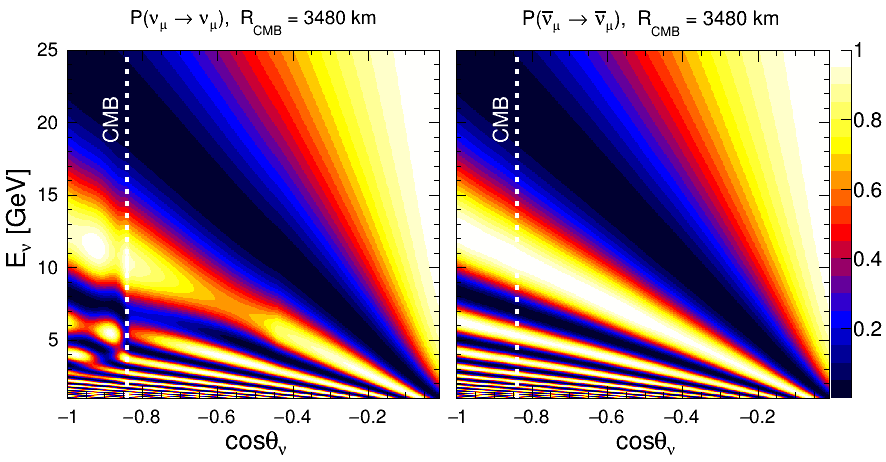}
	\mycaption{Three-flavor $\nu_{\mu}\rightarrow\nu_{\mu}$ oscillograms for the three-layered profile of Earth with the standard $R_\text{CMB}$ of 3480 km as shown by the vertical dotted-white lines. The left (right) panel is for neutrino (antineutrino). We use the three-flavor neutrino oscillation parameters given in Table~\ref{tab:osc-param-value}, where we assume NO and $\sin^2\theta_{23}$ = 0.5.}
	\label{fig:fixed_core_Oscillograms}
\end{figure}

Since the modifications in neutrino oscillation probabilities depend upon the baselines, let us discuss the effects of modification of $R_\text{CMB}$ on neutrino oscillation probabilities in the two-dimensional plane of energy and direction. First, we consider the standard scenario of $R_{\text{CMB}}=3480$ km in Fig.~\ref{fig:fixed_core_Oscillograms}, where we present the three-flavor $\nu_\mu$ survival probability oscillograms in the plane of ($E_\nu$, $\cos\theta_\nu$) for NO. The standard $R_\text{CMB}$ is denoted by a vertical dotted-white line. The left and right panels show the survival probabilities $P(\nu_\mu \rightarrow \nu_\mu)$ and $P({\bar{\nu}_\mu \rightarrow \bar{\nu}_\mu})$, respectively. In each panel, we consider the neutrino energy range of 1 to 25 GeV and neutrino zenith angle ($\cos\theta_\nu$) range of -1 to 0. Here, $\cos\theta_\nu$= -1 ($\cos\theta_\nu$ = 1) represents the upward-going (downward-going) neutrinos. The dark-blue diagonal band which starts from ($E_\nu$ = 1 GeV, $\cos\theta_\nu$ = 0) and ends at ($E_\nu$ = 25 GeV, $\cos\theta_\nu$ = -1), corresponds to the first oscillation minimum, which is also known as ``oscillation valley''~\cite{Kumar:2020wgz,Kumar:2021lrn}. In the left panel,  the red patch around -0.8 $< \cos\theta_\nu <$ -0.5 and 6 GeV $< E_\nu <$ 10 GeV corresponds to the MSW resonance whereas the yellow patches around $\cos\theta_\nu<$ -0.8 and 3 GeV $< E_\nu <$ 6 GeV are found to be due to the NOLR/ parametric resonance. With NO, both these resonances are experienced only by neutrinos. As far as antineutrinos are concerned, we do not see the MSW and the NOLR/ parametric resonance in the right panel for normal ordering. On the other hand, for inverted ordering, this trend is opposite, i.e., antineutrinos experience a significant amount of matter effects whereas neutrinos do not. In this section, we will only focus on neutrino survival channel and NO. 

\begin{figure}
	\centering
	
	\includegraphics[width=0.95\linewidth]{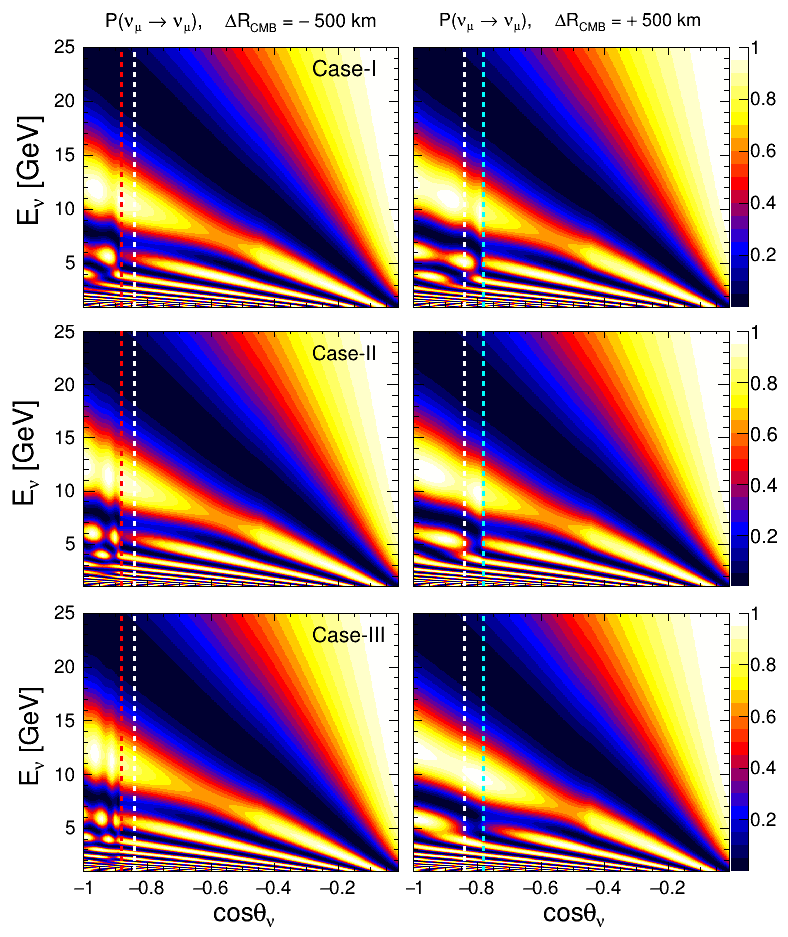}
	\mycaption{Three-flavor $\nu_{\mu}\rightarrow\nu_{\mu}$ oscillograms for the three-layered profile of Earth with modified $R_\text{CMB}$. The left (right) panel is for SC (LC) with $\Delta R_\text{CMB} = -500$ km ($+500$ km). The white, red, and cyan vertical dotted lines correspond to the standard, smaller, and larger $R_\text{CMB}$, respectively. The top, middle and bottom panels correspond to the Case-I, Case-II and Case-III, respectively. We use three-flavor neutrino oscillation parameters given in Table~\ref{tab:osc-param-value}, where we assume NO and $\sin^2\theta_{23}$ = 0.5.}
	\label{fig:Oscillograms_case_1}
\end{figure}

Now, we describe how do the neutrino oscillograms modify during the modification of CMB radius. The top, middle, and bottom rows in Fig.~\ref{fig:Oscillograms_case_1} present the $\nu_{\mu}$ survival probability oscillograms for the Case-I, Case-II, and Case-III, respectively. The left (right) panels correspond to the SC (LC) scenario. The observations are as follows:
\begin{itemize}
	\item Comparing the oscillograms in Fig.~\ref{fig:Oscillograms_case_1} with those in Fig.~\ref{fig:fixed_core_Oscillograms}, we observe that, for both Case-I and Case-II, significant effect on the probabilities occurs in the NOLR/ parametric resonance region.
	
	\item In the standard core scenario (left panel of Fig.~\ref{fig:fixed_core_Oscillograms}), there are three yellow patches in the NOLR/ parametric resonance region. In both Case-I and Case-II, the SC scenario shows that one of the yellow patch is missing and the other two are shrunk. In the LC scenario,  these patches are stretched towards lower baselines.
	
	\item However, for Case-III, we see that the NOLR/ parametric resonance as well as the MSW resonance region both are affected in the SC and LC scenarios, the effect being more prominent for LC.
\end{itemize}

\begin{figure}
	\centering
	
	\includegraphics[width=0.88\linewidth]{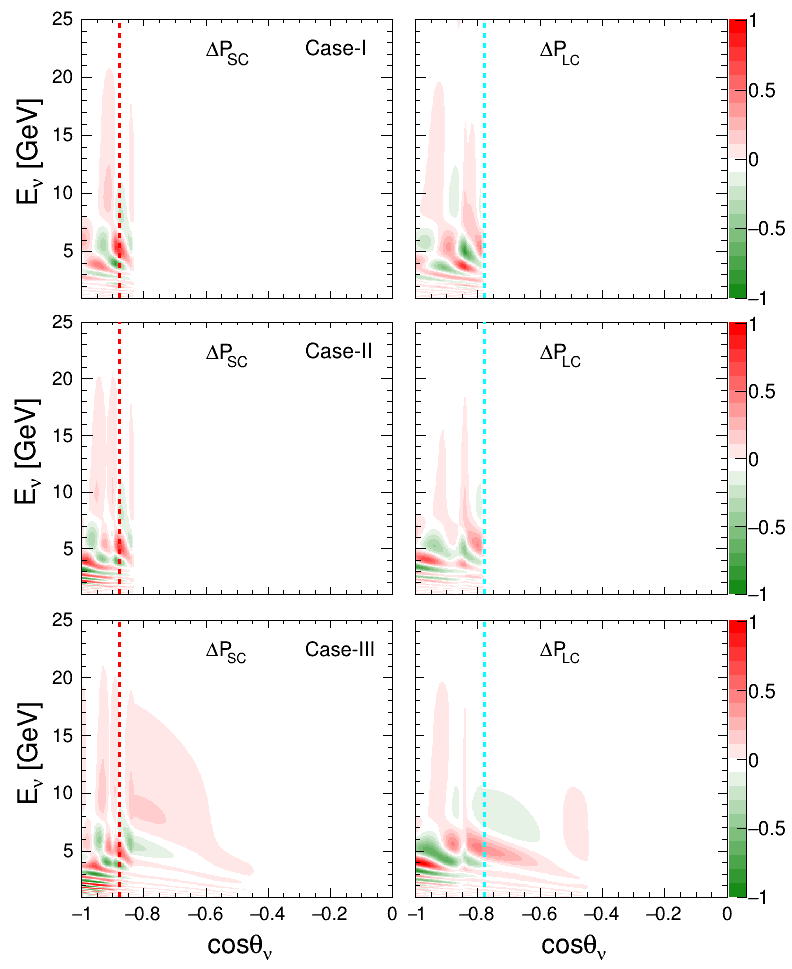}
	
	\mycaption{Three-flavor $\nu_{\mu}\rightarrow\nu_{\mu}$ survival probability difference oscillograms for the three-layered profile of Earth. Here, $\Delta P_{\text{SC}}$ ($\Delta P_{\text{LC}}$)  in the left (right) panels denote P($\nu_{\mu}\rightarrow\nu_{\mu}$) survival probability difference between the standard core and SC (LC) scenario with $\Delta R_\text{CMB} = -500$ km ($+500$ km). The dotted-red and dotted-cyan curves represent the CMB radius for the smaller core and larger core, respectively. Top, middle, and bottom rows correspond to the Case-I, Case-II, and Case-III, respectively. We use the three-flavor neutrino oscillation parameters given in Table~\ref{tab:osc-param-value}, where we assume NO and $\sin^2\theta_{23}$ = 0.5. }
	\label{fig:Diff_Oscillograms_case_1}
\end{figure}

For a clear picture of the energies and baselines where the probability is affected significantly, we present Fig.~\ref{fig:Diff_Oscillograms_case_1} which shows the difference between  $\nu_{\mu}$ survival probability for the standard $R_\text{CMB}$  and SC/LC scenario. The left and right panels show the values of $\Delta P_{\text{SC}}$ and $\Delta P_{\text{LC}}$, respectively, which are defined as 
\begin{align}
\Delta P_{\text{SC}} &= P(\nu_{\mu} \rightarrow \nu_\mu)_{\text{standard}} - P(\nu_{\mu} \rightarrow \nu_\mu)_{\text{SC}}\,, \\
\Delta P_{\text{LC}} &= P(\nu_{\mu} \rightarrow \nu_\mu)_{\text{standard}} - P(\nu_{\mu} \rightarrow \nu_\mu)_{\text{LC}}\,.
\end{align} 
It may be observed that the probability differences are nonzero only at the core regions for Case-I (top panels) and Case-II (middle panels). $\Delta P_{\text{SC/LC}}$ vanishes for the mantle region because in both these cases, the density of the mantle remains the same (see Table~\ref{tab:CMB_variation}). In contrast, for the Case-III (bottom panels), the differences are significant in the core as well as mantle.

From Figs.~\ref{fig:fixed_core_Oscillograms}, \ref{fig:Oscillograms_case_1}, and \ref{fig:Diff_Oscillograms_case_1}, we can infer that the effect of $R_\text{CMB}$ modification manifests itself mainly at the higher baselines and lower energies. The dependence of this effect on the zenith angle ($\cos\theta_\nu$) implies that we need a detector like ICAL with high directional resolution to probe the position of the CMB.

\section{Event Generation at ICAL}
\label{sec:events}

The 50 kton magnetized iron calorimeter (ICAL) detector at the proposed India-based Neutrino Observatory (INO)~\cite{ICAL:2015stm} is going to detect the atmospheric neutrinos and antineutrinos separately in multi-GeV energy range and over a wide range of baselines. ICAL, with a total size of 48 m $\times$ 16 m $\times$ 14.5 m, will have about 151 alternative layers of iron having a thickness of 5.6 cm, with glass Resistive Plate Chambers (RPCs)~\cite{SANTONICO1981377,Bheesette:2009yrp,Bhuyan:2012zzc} sandwiched between the iron layers. The iron plates act as passive detector elements for the neutrino interactions, whereas RPCs act as active detector elements. In other words, the iron plates provide the target mass for neutrino interactions, and RPCs detect the secondary particles like muons and hadrons that are produced during the charged-current (CC) interactions of neutrinos with the iron nuclei. The charged particles deposit their energies in the form of hits in the RPCs during their propagation inside the detector.  The X and Y coordinates of the hits are given by the pickup strips using the produced electronic signals. At the same time, the RPC layer number provides the Z coordinate of the hit. 

A muon in the multi-GeV energy range is a minimum ionizing particle. Hence, it can pass through many layers, leaving a hit in each layer. These hits produced by muons form a track-like event. The magnetic field of about 1.5 T~\cite{Behera:2014zca} enables ICAL to distinguish between the atmospheric neutrinos and antineutrinos by identifying the opposite curvature of $\mu^-$ and $\mu^+$ tracks. Further the nanosecond-level time resolution of RPCs~\cite{Dash:2014ifa,Bhatt:2016rek,Gaur:2017uaf} helps ICAL to separately identify the upward-going and downward-going muon events. In the multi-GeV energy range, the resonance and deep inelastic scatterings (DIS) give rise to the production of hadrons which can deposit energy in the form of multiple hits in the same layer of RPC, resulting into shower-like events. During neutrino interactions, a large fraction of neutrino energy is carried away  by the hadrons, which is quantified as ${E'}_\text{had} = E_\nu - E_\mu$.

In this work, we simulate the unoscillated neutrino events using the NUANCE~\cite{Casper:2002sd} Monte Carlo (MC) neutrino event generator with the ICAL detector geometry as a target. The atmospheric neutrino flux at the proposed INO site~\cite{Athar:2012it,Honda:2015fha} at Theni district of Tamil Nadu, India, is used as an input to NUANCE. The solar modulation effect on atmospheric neutrino flux is incorporated by considering the flux with high solar activity (solar maximum) for half exposure and low solar activity (solar minimum) for another half exposure. A mountain coverage of about 1 km (3800 m water equivalent) at the INO site acts as a filter to reduce the downward-going cosmic muon background by a factor of $\sim 10^6$ ~\cite{Dash:2015blu}. Further, the ICAL analysis considers the events having vertices far from the edges and completely inside the detector to exclude the events entering from outside~\cite{ICAL:2015stm}. Therefore, the background due to the downward-going cosmic muons is expected to be negligible. In our analysis, the statistical uncertainties are minimized by generating MC unoscillated neutrino events for a large exposure of 1000 years for the ICAL detector. The three-flavor neutrino oscillations in the presence of Earth's matter effects are incorporated using the reweighting algorithm~\cite{Devi:2014yaa,Ghosh:2012px,Thakore:2013xqa}.

In the present analysis, we incorporate the detector response for muons and hadrons as described in refs.~\cite{Chatterjee:2014vta,Devi:2013wxa}. These detector responses are obtained by ICAL collaboration after performing a detailed GEANT4~\cite{Geant4:2003} simulation of the ICAL detector~\cite{Chatterjee:2014vta,Devi:2013wxa}. The procedure for obtaining the detector response for muons has been discussed in detail in ref.~\cite{Chatterjee:2014vta}. The muon detector response is simulated by passing a large number of muons through the ICAL detector. The muon forms a track-like event while passing through various RPC layers. These tracks are fitted using a Kalman filter technique to obtain the energy, direction, and charge of reconstructed muons~\cite{Bhattacharya:2015bsp}. The ICAL reconstruction algorithm requires at least 8 to 10 muon hits to reconstruct a track. Since muon deposits an energy of about 100 MeV in each layer of iron, the energy threshold of ICAL is about 1 GeV. Reference~\cite{Chatterjee:2014vta} provides the reconstruction efficiency, energy resolution, angular resolution, and charge identification (CID) efficiency of ICAL for muons (figures 13, 11, 6 and 14 therein, respectively).

Along with the reconstruction of muons, ICAL can also retrieve information about the hadron energy using the total number of hits in the shower-like events. The details about the hadron energy resolution of the ICAL detector is given in ref.~\cite{Devi:2013wxa}. Since the hadron energy resolution is much poorer than the muon energy resolution, we do not add them to obtain neutrino energy. Instead, to exploit the measured four-momentum of muon and hadron energy on an event-by-event basis, we employ a binning scheme having reconstructed muon energy ($E_\mu^\text{rec}$), muon direction ($\cos\theta_\mu^\text{rec}$), and hadron energy (${E^\prime}_{\text{had}}^\text{rec}$) as three independent observables. The detector properties are folded in following the procedure mentioned in~\cite{Ghosh:2012px,Thakore:2013xqa,Devi:2014yaa}. For the analysis, the reconstructed events are scaled from 1000-yr MC to 20-yr MC. For 1 Mt$\cdot$yr exposure of ICAL, we would get about 8850 reconstructed $\mu^-$ and 4032 reconstructed $\mu^+$ events using the three-flavor neutrino oscillation for propagation through the Earth's matter considering the three-layered profile of Earth with the standard core if the mass ordering is normal. In Table~\ref{tab:events}, we present the number of reconstructed $\mu^-$ and $\mu^+$ events for NO for all the three cases for SC and LC scenarios for the exposure of 1 Mt$\cdot$yr at the ICAL detector. It is important to note that the total event rate in Table~\ref{tab:events} is almost identical for all the three cases. However, after binning these events in the above three observables, the three cases would look quite different. This is possible because ICAL would have good energy and directional resolutions for reconstructed muons.

\begin{table}
	\centering
	\begin{tabular}{| c | c | c | c | c |} 
		\hline \hline
			  &  \multicolumn{2}{c|}{$\mu^-$ events} & 
			\multicolumn{2}{c|}{$\mu^+$ events}  \\
			\cline{2-5}
			&  SC & LC & SC & LC \\ 
			\hline 
			
			Case-I & 8842 & 8858 & 4032 & 4032 \\
			
			\hline
			
			Case-II & 8844 & 8862 & 4030 & 4034 \\ 
		   
		    \hline
		    
		    Case-III & 8850 & 8876 & 4032 & 4032 \\ 
			
			\hline \hline
	\end{tabular}
	\mycaption{The total number of reconstructed $\mu^-$ and $\mu^+$ events expected at the 50 kt ICAL detector in 20 years for the cases I, II, and III. The SC (LC) scenario corresponds to  $\Delta R_\text{CMB} = -500$ km ($+500$ km). The number of reconstructed events for the standard CMB is 8850 (4032) for $\mu^-$ ($\mu^+$). We use the three-flavor neutrino oscillation parameters given in Table~\ref{tab:osc-param-value}, where we assume NO and $\sin^2\theta_{23}$ = 0.5.
	}
	\label{tab:events}
\end{table}

\begin{figure}
	\centering
	\includegraphics[width=0.95\linewidth]{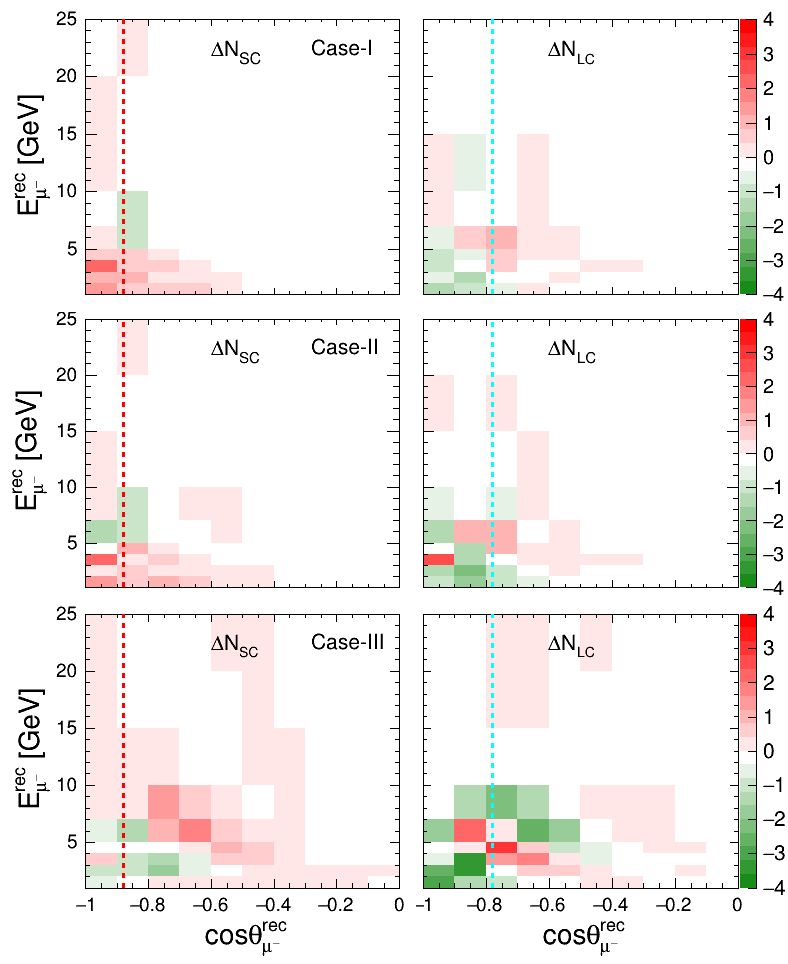}
	\mycaption{The ($E{^\text{rec}_{\mu^-}}$, $\cos\theta{^\text{rec}_{\mu^-}}$) distributions of the differences in reconstructed $\mu^-$ events with the standard and modified $R_\text{CMB}$. We have taken an exposure of 1 Mt$\cdot$yr at the ICAL detector. In the left (right) panels, $\Delta N_{\text{SC}}$ ($\Delta N_{\text{LC}}$) denote the $\mu^-$ event differences between the standard core and SC (LC) scenario with $\Delta R_\text{CMB} = -500$ km ($+500$ km). The dotted-red and dotted-cyan curves represent the CMB radius for the smaller core and larger core, respectively. Top, middle, and bottom rows correspond to the Case-I, Case-II, and Case-III, respectively. We use the three-flavor neutrino oscillation parameters given in Table~\ref{tab:osc-param-value}, where we assume NO and $\sin^2\theta_{23}$ = 0.5.}
	\label{fig:Event_dif}
\end{figure}

Now we discuss the effect of $R_\text{CMB}$ modification on the distributions of reconstructed muon events at the ICAL detector. In the left (right) panels of Fig.~\ref{fig:Event_dif}, we present the distributions of differences between events of the standard core and that of smaller (larger) core in the plane of ($E{^\text{rec}_{\mu^-}}$, $\cos\theta{^\text{rec}_{\mu^-}}$) for 20 years (1 Mt$\cdot$yr) of exposure considering NO. For demonstrating the difference of reconstructed event distributions with 1 Mt$\cdot$yr exposure, we use a binning scheme\footnote{For $E_\mu^\text{rec}$, we take 5 bins of 1 GeV in $(1 - 5)$ GeV, 1 bin of 2 GeV in $(5 - 7)$ GeV, 1 bin of 3 GeV in $(7 - 10)$ GeV, and 3 bins of 5 GeV in $(10 - 25)$ GeV. On the other hand, for  $\cos\theta_\mu^\text{rec}$, we choose uniform bins with a width of 0.1 in the range of $-1$ to $1$. Note that for analysis, we will use a finer binning scheme as described in Sec.~\ref{sec:binning_scheme}.} where we have 10 bins in $E_\mu^\text{rec}$ and 20 bins in $\cos\theta_\mu^\text{rec}$. Since for NO, significant amount of matter effects are present for neutrinos but not antineutrinos, we choose to present the distributions of event difference in Fig.~\ref{fig:Event_dif} for $\mu^-$ only. The event difference is quite small for $\mu^+$, and hence, not shown here. For Case-I and Case-II, event differences mainly occur in the core region, although some differences can also be observed in the mantle region due to the angular smearing originating from the difference between the directions of incoming neutrinos and reconstructed muons. On the other hand, for Case-III, we see significant event differences which span both the core as well as in the mantle regions. In the next section, we describe the numerical procedure used to estimate the median sensitivity of ICAL to probe the CMB.

\section{The Analysis Method}
\label{sec:statistical analysis}

\subsection{Binning Scheme for Analysis}
\label{sec:binning_scheme}
In this section, we present the binning scheme used for numerical analysis. From Figs.~\ref{fig:Diff_Oscillograms_case_1} and \ref{fig:Event_dif}, it is clear that $R_\text{CMB}$ modification manifests itself mainly through the regions of higher baselines ($-1<\cos\theta<-0.7$) and lower energies ($1<E_\nu<6$ GeV). Therefore, we use finer bins for this region. The optimized binning scheme is shown in Table~\ref{tab:binning_scheme}. We have total 16 bins for $E^{\text{rec}}_\mu$ spanning in the range of 1 to 25 GeV, 39 bins for $\cos\theta^{\text{rec}}_\mu$ from -1 to 1, and 4 bins for $E^{\prime \text{rec}}_{\text{had}}$ in the range of 0 to 25 GeV. For our analysis, only upward-going neutrinos are relevant because they have experienced the matter effects, but we have also included downward-going neutrinos ($0<\cos\theta <1$)  because they help in reducing the impact of normalization errors in atmospheric neutrino events. This also includes those upward-going (near horizon) neutrino events that result in the downward-going reconstructed muon events due to angular smearing during the interaction of neutrinos as well as reconstruction. We have considered the same binning scheme for $\mu^-$ and $\mu^+$. 

\begin{table}
	\centering
	\begin{tabular}{|c|c|c|c c|}
		\hline 
		Observable & Range & Bin width & \multicolumn{2}{c|}{Number of bins} \\ 
		\hline 
		\multirow{4}{*}{$E_\mu^{\rm rec}$ (GeV)}  & [1, 6] & 0.5 & 10& 
		\rdelim\}{4}{7mm}[16] \cr
		& [6, 12] & 2 & 3 &\cr
		& [12, 15] & 3 & 1 & \cr
		& [15, 25] & 5 & 2 & \cr
		\hline
		\multirow{4}{*}{$\cos\theta^{\text{rec}}_\mu$} & [-1.0, -0.85] & 0.0125 & 12 & \rdelim\}{4}{7mm}[39]\cr
		& [-0.85, -0.4] & 0.025 & 18 &\cr
		& [-0.4, 0] & 0.1 & 4 & \cr
		& [0, 1] & 0.2 & 5 & \cr
		\hline
		\multirow{3}{*}{$E^{\prime \text{rec}}_{\text{had}}$ (GeV)}  & [0, 2] & 1 & 2  & \rdelim\}{3}{7mm}[4] \cr
		& [2, 4] & 2 & 1 &\cr
		& [4, 25] & 21 & 1 &\cr
		\hline 
		
	\end{tabular}
	\mycaption{The optimized binning scheme considered for the numerical analysis of the present work for reconstructed observables $E_\mu^{\rm rec}$, $\cos\theta^{\text{rec}}_\mu$, and $E^{\prime \text{rec}}_{\text{had}}$ for both reconstructed $\mu^-$ and $\mu^+$.}
	\label{tab:binning_scheme}
\end{table}

\subsection{Numerical Analysis}

To estimate the sensitivity of ICAL, a $\chi^2$ analysis is performed which is expected to give median sensitivity in the frequentist approach~\cite{Blennow:2013oma}. For this analysis, we define the following Poissonian $\chi^2_-$~\cite{Baker:1983tu} for reconstructed $\mu^-$ events as considered in ref.~\cite{Devi:2014yaa}:
\begin{equation}\label{eq:chisq_mu-}
\chi^2_- = \mathop{\text{min}}_{\xi_l} \sum_{i=1}^{N_{{E'}_\text{had}^\text{rec}}} \sum_{j=1}^{N_{E_{\mu}^\text{rec}}} \sum_{k=1}^{N_{\cos\theta_\mu^\text{rec}}} \left[2(N_{ijk}^\text{theory} - N_{ijk}^\text{data}) -2 N_{ijk}^\text{data} \ln\left(\frac{N_{ijk}^\text{theory} }{N_{ijk}^\text{data}}\right)\right] + \sum_{l = 1}^5 \xi_l^2\,,
\end{equation}
with 
\begin{equation}
N_{ijk}^\text{theory} = N_{ijk}^0\left(1 + \sum_{l=1}^5 \pi^l_{ijk}\xi_l\right)\,.
\label{eq:chisq_2}
\end{equation}
Here $N_{ijk}^\text{theory}$ and $N_{ijk}^\text{data}$ correspond to the expected and observed number of reconstructed $\mu^-$ events in a given ($E^{\text{rec}}_\mu$, $\cos\theta^{\text{rec}}_\mu$, $E^{\prime \text{rec}}_{\text{had}}$) bin, respectively. The quantity $N_{ijk}^0$ stands for the number of expected events without considering systematic uncertainties. From the binning scheme mentioned in Table~\ref{tab:binning_scheme}, $N_{E_{\mu}^\text{rec}}$ = 16, $N_{\cos\theta_\mu^\text{rec}}$ = 39, and $N_{{E'}_\text{had}^\text{rec}}$ = 4. For our analysis, we use the well-known method of pulls~\cite{Gonzalez-Garcia:2004pka,Huber:2002mx,Fogli:2002pt} to incorporate the following five systematic uncertainties~\cite{Ghosh:2012px,Thakore:2013xqa}: (i) 20\% uncertainty on flux normalization, (ii) 10\% uncertainty on cross section, (iii) 5\% energy dependent tilt error in flux, (iv) 5\% zenith angle dependent tilt error in flux, and (v) 5\% overall systematics. The pull variables of systematic uncertainties are denoted by $\xi_l$ in Eqs.~\ref{eq:chisq_mu-} and ~\ref{eq:chisq_2}. 

The $\chi^2_+$ for reconstructed $\mu^+$ events is also obtained by following the same procedure. The separate contributions of both $\chi^2_-$ and $\chi^2_+$ are added to get the resultant sensitivity of the ICAL detector, which is define as $\chi^2$:
\begin{equation}
\chi^2 = \chi^2_- + \chi^2_+\,.
\end{equation}
To simulate the MC data for our analysis, we use the benchmark values of oscillation parameters given in Table~\ref{tab:osc-param-value} as true parameters. In the fit, we minimize $\chi^2$ over the pull variables $\xi_l$  and the relevant oscillation parameters. We vary the atmospheric mixing angle $\sin^2\theta_{23}$ in the range $(0.36 - 0.66)$ and atmospheric mass-squared difference $|\Delta m^2_{\text{eff}}|$ in the range $(2.1 - 2.6)\times 10^{-3}$ eV$^2$. We also minimize over both the choices of neutrino mass orderings, NO and IO. During the fit, we do not vary solar oscillation parameters $\sin^2 2\theta_{12}$ and $\Delta m^2_{21}$; they are kept fixed at their true values given in Table~\ref{tab:osc-param-value}. For the reactor mixing angle, which is already very well measured~\cite{Capozzi:2021fjo,NuFIT,Esteban:2020cvm,deSalas:2020pgw}, we take a fixed value of $\sin^2 2\theta_{13}$ = 0.0875 both in MC data and theory. We kept $\delta_{\rm CP}$ = 0 throughout our analysis in both data and theory.

\section{Results}
\label{sec:results}

In this section, we present the statistical significance of the ICAL detector to probe the location of the core-mantle boundary. For numerical analysis, we simulate the MC data assuming the three-layered profile as a true profile of the Earth with standard core. We quantify the statistical significance of the analysis for measuring the position of CMB in the following way:
\begin{align}
\Delta \chi^2_{\text{CMB}} = \chi^2(\text{modified } R_\text{CMB}) - \chi^2 (\text{standard } R_\text{CMB})\,,
\label{eq:delta_chisq}
\end{align}
where $\chi^2$(modified $R_\text{CMB}$) and $\chi^2$(standard $R_\text{CMB}$) is calculated by performing a fit to the MC data with the modified and standard $R_\text{CMB}$, respectively. Here, we calculate the Asimov sensitivity representing the median $\Delta \chi^2_{\text{CMB}}$ in the frequentist approach where the statistical fluctuations are suppressed such that $\chi^2$(standard $R_\text{CMB}$) $\approx$ 0.

\begin{table}
	\centering
	\begin{tabular}{|cc|c|c|c|c|}
			\hline
			\multirow{2}{*}{} & \multirow{2}{*}{} &  \multicolumn{2}{c|}{500 kt$\cdot$yr} & 
			\multicolumn{2}{c|}{1 Mt$\cdot$yr}  \\
			\cline{3-6}
			& & w/ CID  & w/o CID & w/ CID  & w/o CID  \\ 
			\hline 
			
			\multirow{2}{*}{Case-I} & SC & 0.76 (0.76) & 0.50 (0.51) & 1.53 (1.53) &  1.01 (1.02) \\
			
			& LC &  0.72 (0.72) &  0.47 (0.48) &  1.44 (1.44) &  0.95 (0.95) \\
			\hline 
			\multirow{2}{*}{Case-II} & SC &  1.27 (1.30) &  0.83 (0.84) & 2.53 (2.59) & 1.66 (1.68)\\
			
			& LC &  1.33 (1.33) & 0.84 (0.84) & 2.63 (2.65) &  1.67 (1.69) \\ 
			\hline 
			\multirow{2}{*}{Case-III} & SC &  1.04 (1.06) &  0.66 (0.67) &  2.09 (2.12) &  1.33 (1.35) \\ 
			
			& LC &  4.06 (4.18) & 2.63 (2.69) & 8.07 (8.34) & 5.23 (5.37)\\
			
			\hline
	\end{tabular}
	\mycaption{$\Delta \chi^2_{\text{CMB}}$ sensitivities for the modification of $R_\text{CMB}$ by $\pm\,$500 km for all three cases with an exposure of 500 kt$\cdot$yr (10 years) and 1 Mt$\cdot$yr (20 years). For the numbers without parentheses, the $\Delta \chi^2_{\text{CMB}}$  minimization has been performed by varying test values of $\sin^2\theta_{23}$ and  $\Delta m^2_{\text{eff}}$ over their current uncertainties, and taking both mass orderings NO and IO. The numbers given in parentheses correspond to a fixed-parameter scenario where we do not marginalize over oscillation parameters in the fit. The results in the second and fourth (third and fifth) columns are with (without) charge identification. For simulating MC data, we use three-flavor neutrino oscillation parameters given in Table~\ref{tab:osc-param-value}, where we assume NO and $\sin^2\theta_{23}$ = 0.5.}
	\label{tab:results}
\end{table}

In Table~\ref{tab:results}, we present the $\Delta \chi^2_{\text{CMB}}$ sensitivity with an exposure of 500 kt$\cdot$yr and 1 Mt$\cdot$yr for all three cases in the SC ($R_{\text{CMB}}$ = 2980 km) and LC ($R_{\text{CMB}}$ = 3980 km) scenarios. It is apparent from Table~\ref{tab:results} that the ICAL detector is sensitive to modification of $R_\text{CMB}$ by $\pm\,$500 km with a statistical significance of more than 1$\sigma$ for all the three cases with an exposure of 1 Mt$\cdot$yr and the CID capability of ICAL detector plays a crucial role in achieving these sensitivities. Comparing the numbers which are given with and without parentheses in Table~\ref{tab:results}, we learn that the impact of marginalization over oscillation parameters in the fit is negligible in our study.

\begin{figure}
	\centering
	\includegraphics[width=0.85\linewidth]{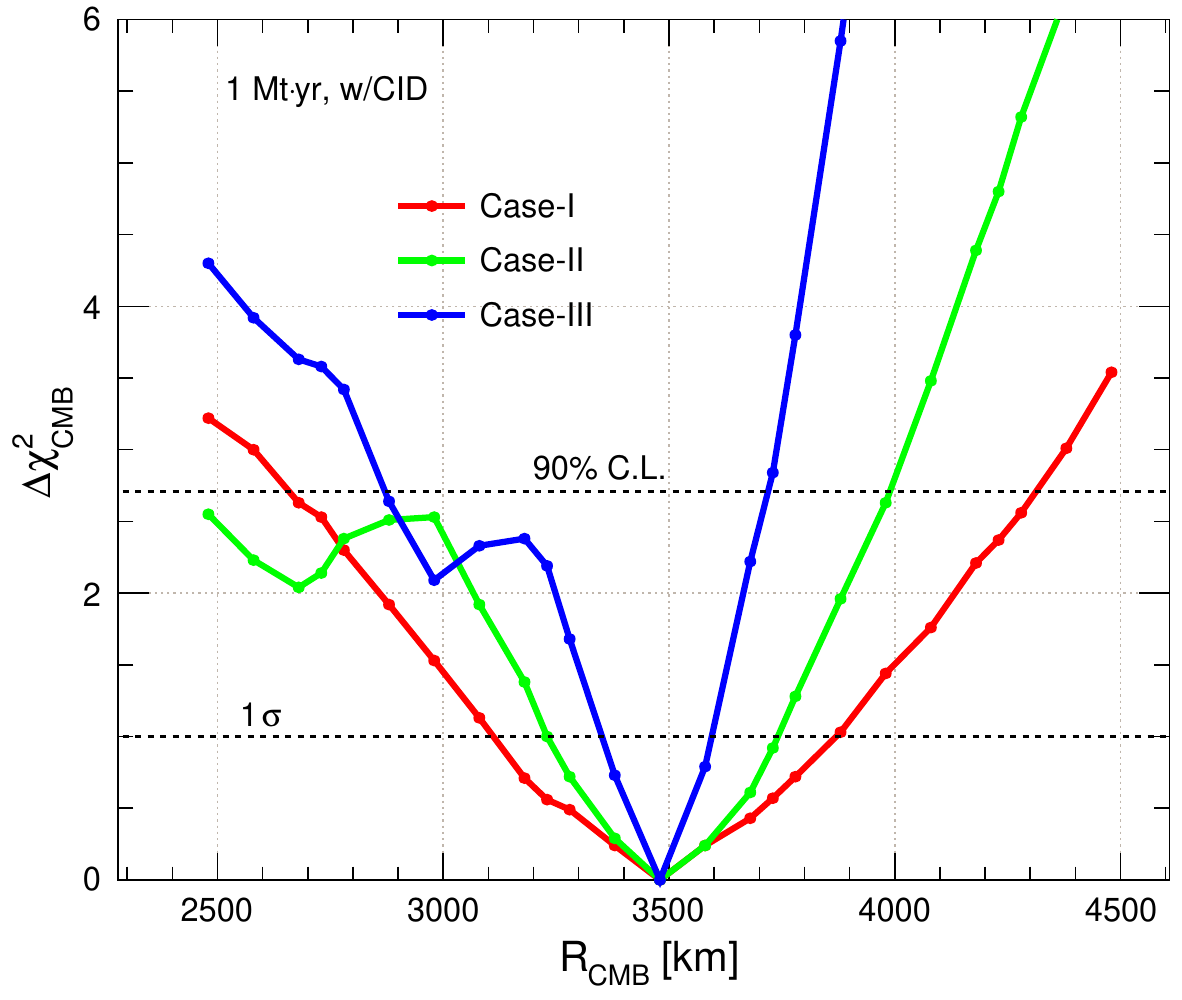}
	\mycaption{The median $\Delta \chi^2_{\text{CMB}}$ sensitivities as functions of the location of CMB. The red, green, and blue curves represent Case-I, Case-II, and Case-III, respectively. The $\Delta \chi^2_{\text{CMB}}$  minimization has been performed by varying test values of $\sin^2\theta_{23}$ and  $\Delta m^2_{\text{eff}}$ over their current uncertainties, and taking both mass orderings NO and IO. The results correspond to 1 Mt$\cdot$yr exposure with charge identification capability of ICAL. For simulating MC data, we use three-flavor neutrino oscillation parameters given in Table~\ref{tab:osc-param-value}, where we assume NO and $\sin^2\theta_{23}$ = 0.5.	
	}
	\label{fig:CMB_variation_chisq_margin}
\end{figure}

In Fig.~\ref{fig:CMB_variation_chisq_margin}, we present the sensitivities of the ICAL detector while modifying $R_\text{CMB}$ in smaller steps of 50 - 100 km up to $\Delta R_\text{CMB} = \pm\,$1000 km with respect to the standard $R_{\text{CMB}}$ of 3480 km. The figure shows that the ICAL detector would be able to measure the location of CMB at 1$\sigma$ with a precision of about $\pm\,$380 km, $\pm\,$250 km, and $\pm\,$120 km for Case-I, Case-II, and Case-III, respectively. We can observe that the sensitivities are in general increasing as we go from Case-I to Case-II and Case-III. Note that Case-II is the most realistic one as mentioned earlier. Needless to mention that the CID capability of ICAL plays a crucial role to achieve this precision in locating $R_\text{CMB}$. For an example, in the absence of CID, the 1$\sigma$ precision for Case-II would be around $\pm\,330$ km. 
 
The most striking feature is the asymmetry and non-monotonic behavior observed in $\Delta \chi^2_\text{CMB}$ about the standard $R_\text{CMB}$ value in Case-II and Case-III. This turns out to be the effect of NOLR/ parametric resonance as explained in appendix~\ref{app:Case-II-variation}. We also evaluate the sensitivity for the 81-layered PREM profile for Case-II in appendix~\ref{app:Case-II-prem81}. We find that the $1\sigma$ precision on $R_\text{CMB}$ is about $\pm\,350$ km.

\section{Summary and Concluding Remarks}
\label{sec:conclusion}

Understanding the detailed internal structure of Earth is an active field of research. This quest has been pursued traditionally using gravitational measurements and seismic studies. In recent times, neutrinos have emerged as important messengers to achieve this goal of multi-messenger tomography of Earth. For example, geoneutrinos can shed light on the composition and energy budget of Earth. The oscillations of atmospheric neutrinos at GeV energies and their absorption at TeV-PeV energies deep inside the Earth can provide complementary information on density profile and composition of Earth. We start by discussing  how the information about the interior of Earth is obtained indirectly using gravitational measurements and seismic studies, and focus on how neutrinos can provide complementary information through their weak interactions.
 
While passing through different regions inside Earth, the multi-GeV atmospheric neutrinos experience the Earth's matter effects due to interactions with ambient electrons. These matter effects depend upon the density of electrons and alter the neutrino oscillation probabilities. In particular, neutrinos with energies 6--10 GeV experience MSW resonance while passing through mantle. The core-passing neutrinos with energies 3--6 GeV may also experience the neutrino parametric/ oscillation length resonance (NOLR) which depends upon the density jump at the core-mantle boundary (CMB) and its location. These effects make the neutrino signal at atmospheric neutrino detectors sensitive to the density profile of Earth. In the present work, we explore the effect of changing the radius of the CMB with respect to its standard value, $R_\text{CMB} = 3480$ km, on the neutrino oscillation probabilities, and calculate the sensitivity of the ICAL detector at INO for measuring the CMB location.

We perform our analysis in the approximation of three-layered profile of Earth:  core, inner mantle and outer mantle. While exploring the effect of changing CMB location, we consider the modification of the density profile of Earth in three ways. In Case-I, we modify $R_\text{CMB}$ while keeping the density in each layer constant; the total mass of Earth is not constrained. In Case-II, we modify $R_\text{CMB}$ and allow the density of core to modify such that the Earth's mass remains the same. In Case-III, we modify  $R_\text{CMB}$ and the densities of core and inner mantle such that the masses of core and inner mantle, and hence that of Earth, are not affected. For each case, we consider two scenarios: SC (smaller core with $\Delta R_\text{CMB}=-500$ km) and LC (larger core  with $\Delta R_\text{CMB}=+500$ km). For these scenarios, we observe that the effect of modification in $R_\text{CMB}$ manifests itself mainly for core-passing neutrinos in the NOLR/parametric resonance region. Moreover, for Case-III, these effects can also be seen in the MSW resonance region. The good direction and energy resolution of ICAL enables it to preserve these features at the level of reconstructed muons. 

We estimate the sensitivity of ICAL for measuring the position of CMB in the three cases described above. To determine the precision on the location of CMB, we simulate the prospective data with the standard $R_\text{CMB}$ and calculate  $\Delta \chi^2$ when the data are fitted with modified $R_\text{CMB}$ values. We observe that the sensitivity is significantly enhanced by the charge identification capability of ICAL, but not affected much by the uncertainties in oscillation parameters. The ICAL detector would be able to measure the location of CMB at 1$\sigma$ with a precision of about $\pm\,$380 km, $\pm\,$250 km, and $\pm\,$120 km for Case-I, Case-II, and Case-III, respectively. For Case-II, we also perform the sensitivity study for locating CMB using the 81-layered PREM  profile, and find the 1$\sigma$ precision of about $\pm\,$350 km. Note that Case-II is the most realistic one, since it is based on the assumption that the density of mantle is known from seismic studies and the total mass of Earth is well measured from gravitational probes.

Also, we find that the sensitivity is not symmetric about the standard $R_\text{CMB}$ when we go to smaller or larger core radii. At large values of $|\Delta R_\text{CMB}|$ in Case-II and Case-III, the value of $\Delta \chi^2$ is not even monotonic. This a real physical effect, the origin of which may be traced to interesting interference effects in the NOLR/ parametric resonance region. As a result of this, larger core radii would be more constrained as compared to smaller core radii using atmospheric neutrino oscillation data.  

In this study, we have found that the NOLR/ parametric resonance effects play a crucial role in measuring the radius of the CMB. Note that the NOLR/ parametric resonance comes into picture because of the contrast in the density of core and mantle, and specific relation between oscillation phases gained by neutrinos during their travel in mantle and core. Such quantum mechanical effects are possible with neutrino oscillations where details of phase change along the neutrino path are important, but not with neutrino absorption where only the integrated matter profile encountered along the path is relevant. Therefore, neutrino oscillations have a great potential for probing the internal structure of Earth, owing to the current precision of neutrino oscillation parameters.
 
The large amount of data from the next-generation atmospheric neutrino experiments like ORCA, IceCube/DeepCore/Upgrade, Hyper-K, DUNE, and P-ONE will significantly enhance the prospects of neutrino oscillation tomography of Earth. Over time, this data will augment the gravitational measurements and seismic studies to give us a clearer picture of the internal structure of Earth. Since neutrino oscillations are sensitive to the electron number densities, as opposed to the gravitational and seismic measurements that are sensitive to the baryonic number densities and material properties of the Earth's interior, neutrino data will provide independent and complementary information to these traditional probes.  

\subsubsection*{Acknowledgements}

We thank F. Halzen, C. Rott, W. F. McDonough, V. Leki$\acute{c}$, P. Huber, and S. Palomares-Ruiz for useful discussions. We sincerely thank the INO internal referees, Amitava Raychaudhuri and Kamales Kar for their careful reading of the manuscript and for providing useful suggestions. A. Kumar would like to thank the organizers of the ``Multi-messenger Tomography of Earth (MMTE 2022)” workshop at The Cliff Lodge at Snowbird, Salt Lake City, Utah, USA during 30th to 31st July, 2022, for providing him an opportunity to present the preliminary results from this work. We acknowledge the support from the Department of Atomic Energy (DAE), Govt. of India, under the Project Identification Numbers RTI4002 and RIO 4001. S.K.A. is supported by the Young Scientist Research Grant [INSA/SP/YSP/144/2017/1578] from the Indian National Science Academy (INSA). S.K.A. acknowledges the financial support from the Swarnajayanti Fellowship (sanction order No. DST/SJF/PSA- 05/2019-20) provided by the Department of Science and Technology (DST), Govt. of India, and the Research Grant (sanction order No. SB/SJF/2020-21/21) provided by the Science and Engineering Research Board (SERB), Govt. of India, under the Swarnajayanti Fellowship project. S.K.A would like to thank the United States-India Educational Foundation for providing the financial support through the Fulbright-Nehru Academic and Professional Excellence Fellowship (Award No. 2710/F-N APE/2021). A.K.U. acknowledges financial support from the DST, Govt. of India (DST/INSPIRE Fellowship/2019/IF190755). The numerical simulations are carried out using the ``SAMKHYA: High-Performance Computing Facility” at the Institute of Physics, Bhubaneswar, India.

\begin{appendix}

\section{Asymmetric Effects with Smaller and Larger Core}
\label{app:Case-II-variation}

\begin{figure}
	\centering
	\includegraphics[width=0.49\linewidth]{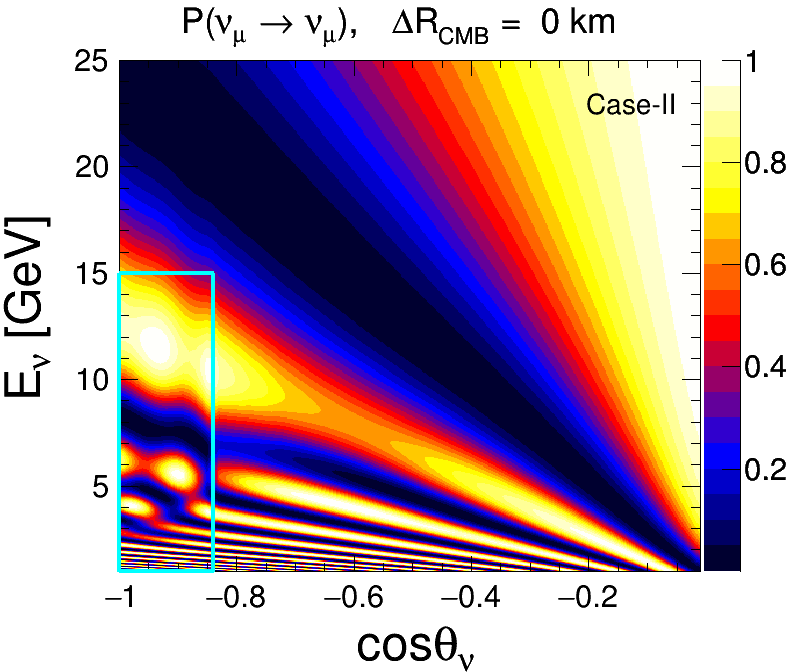}
	\includegraphics[width=0.49\linewidth]{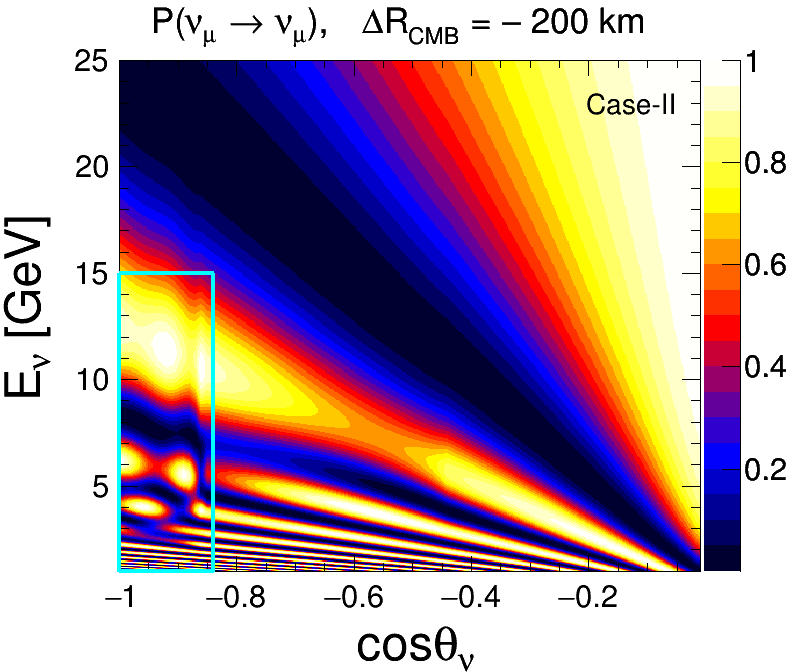} \\ \vspace{2mm}
	\includegraphics[width=0.49\linewidth]{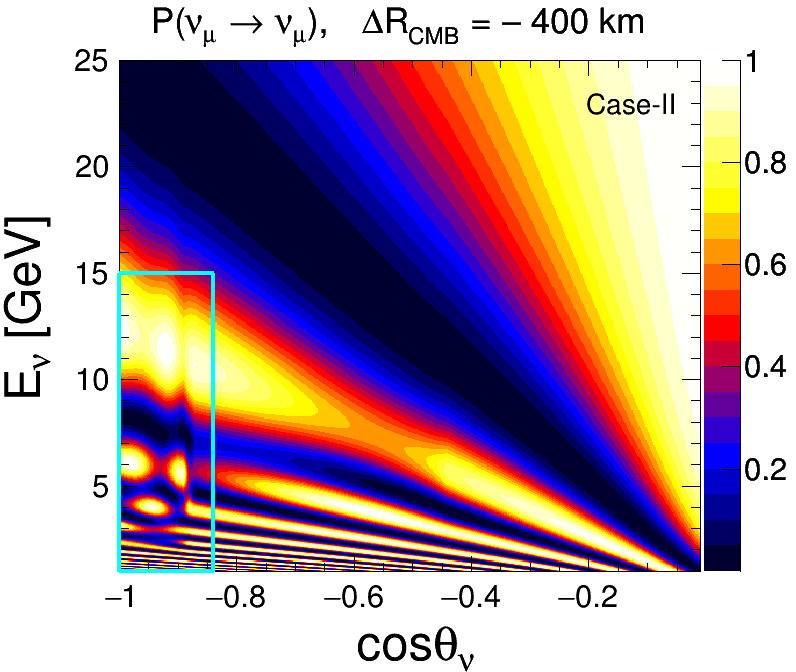}
	\includegraphics[width=0.49\linewidth]{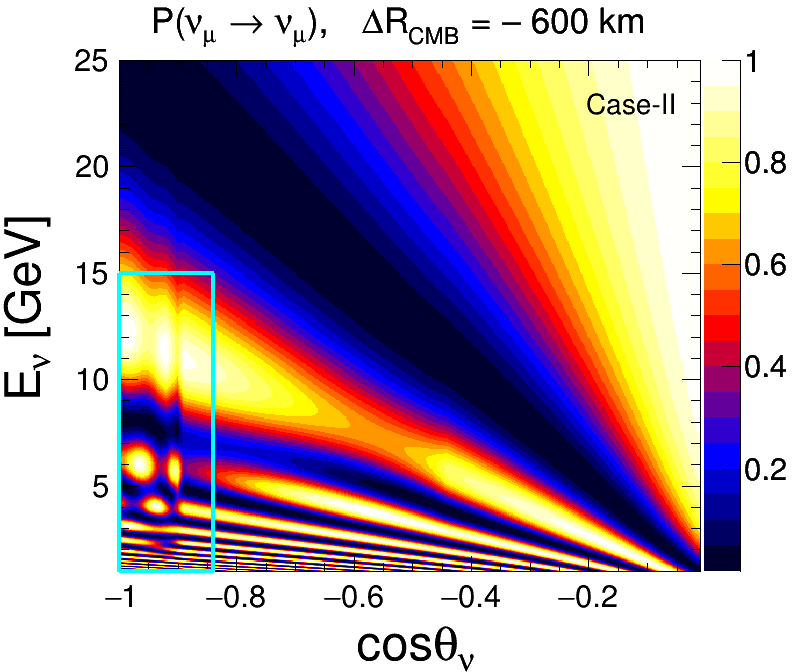} \\ \vspace{2mm}
	\includegraphics[width=0.49\linewidth]{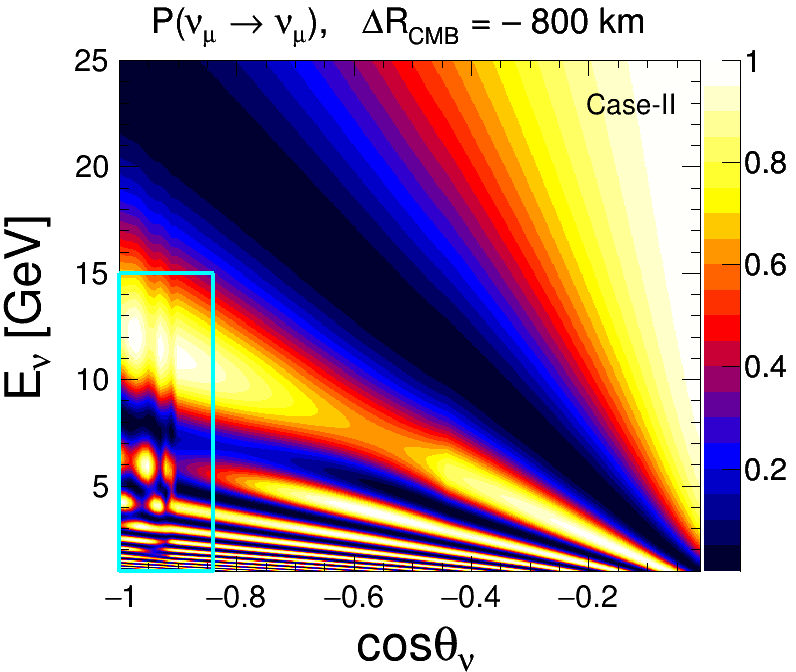}
	\includegraphics[width=0.49\linewidth]{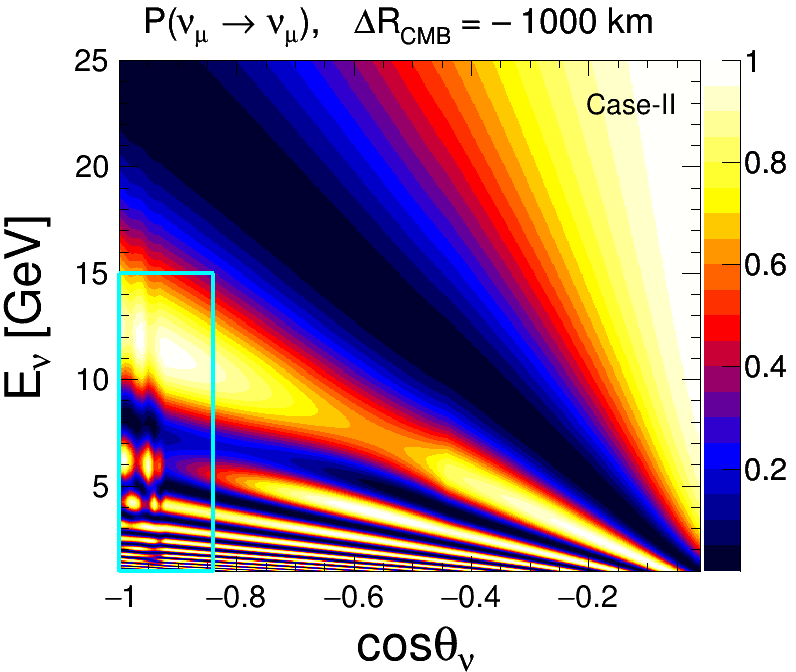}
	\mycaption{Possible impact of smaller core radii on three-flavor $\nu_{\mu}\rightarrow\nu_{\mu}$ oscillograms for the three-layered profile of Earth. The oscillograms are shown for $R_\text{CMB} = 3480 \text{ km} + \Delta R_\text{CMB}$ where $\Delta R_\text{CMB} =  (0, -200, -400, -600, -800, -1000)$ km, with the density profile modified according to Case-II. We use three-flavor neutrino oscillation parameters given in Table~\ref{tab:osc-param-value}, where we assume NO and $\sin^2\theta_{23}$ = 0.5.}
	\label{fig:CMB_varying_case_II_sc}
\end{figure}

\begin{figure}
	\centering
	\includegraphics[width=0.49\linewidth]{images/Oscillograms-appendix/Puu_CMB_fixed_core}
	\includegraphics[width=0.49\linewidth]{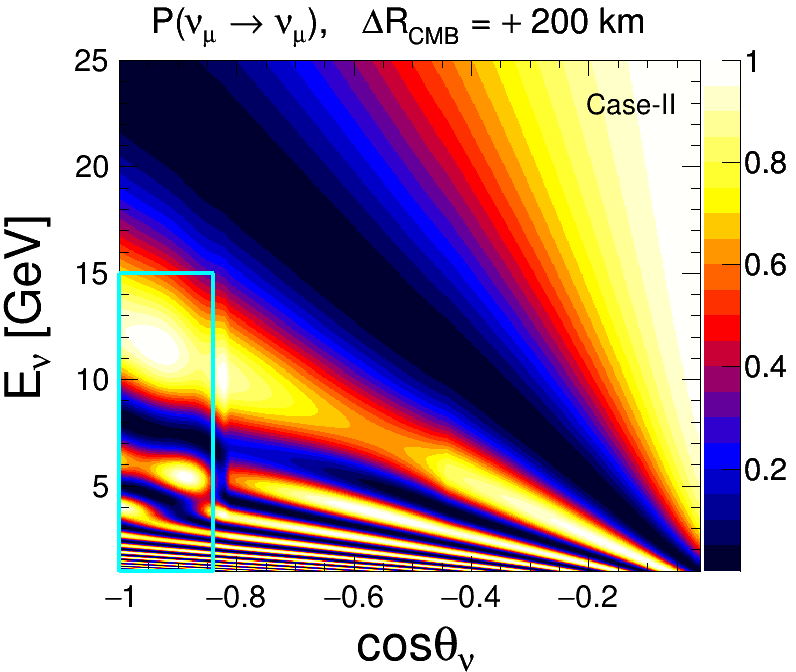} \\ \vspace{2mm}
	\includegraphics[width=0.49\linewidth]{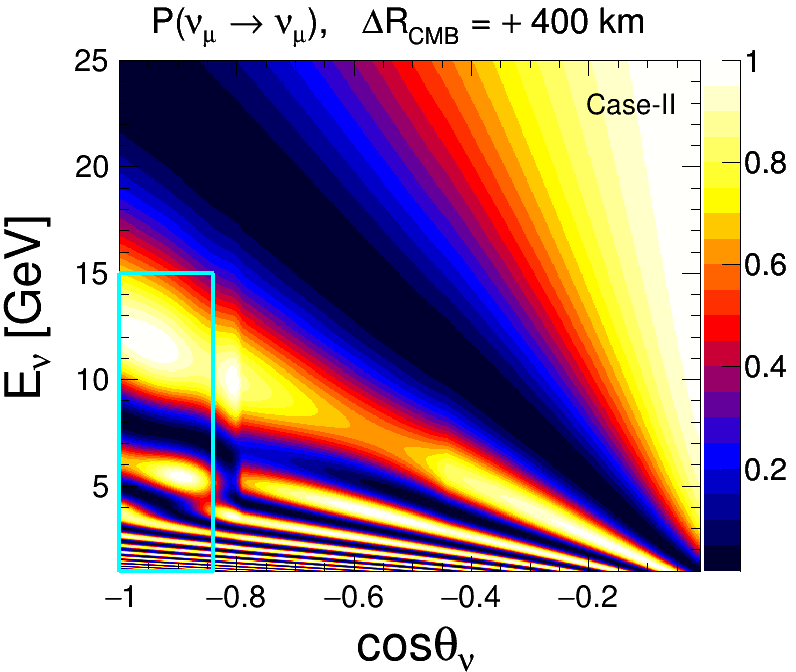}
	\includegraphics[width=0.49\linewidth]{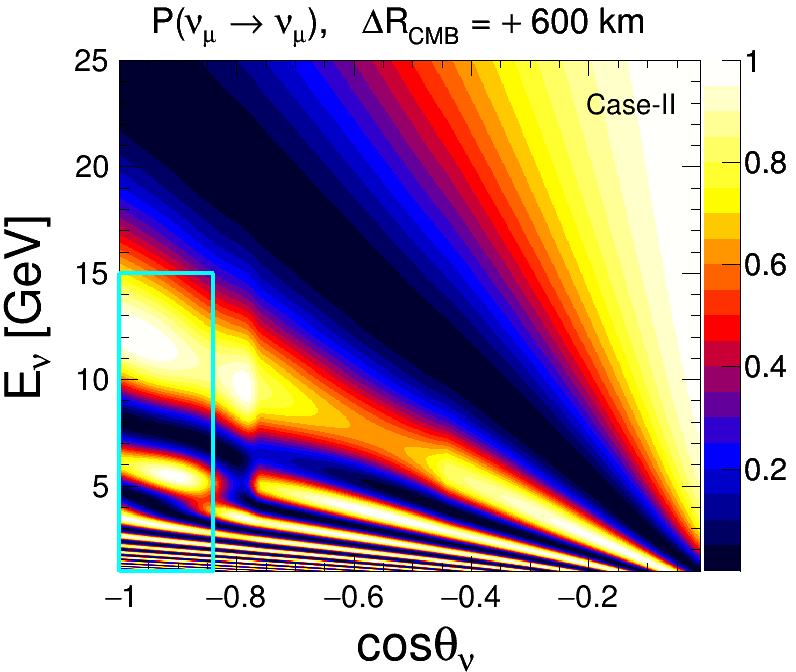} \\ \vspace{2mm}
	\includegraphics[width=0.49\linewidth]{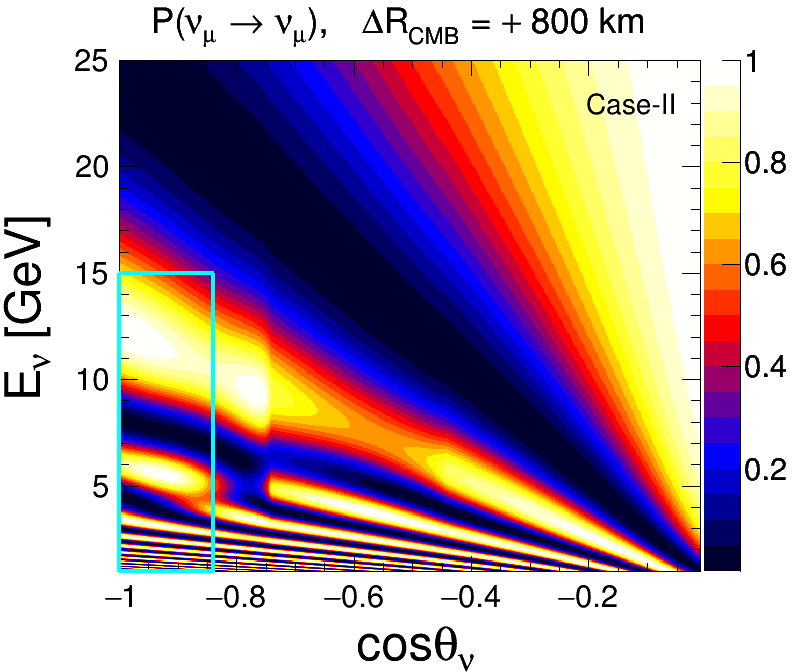}
	\includegraphics[width=0.49\linewidth]{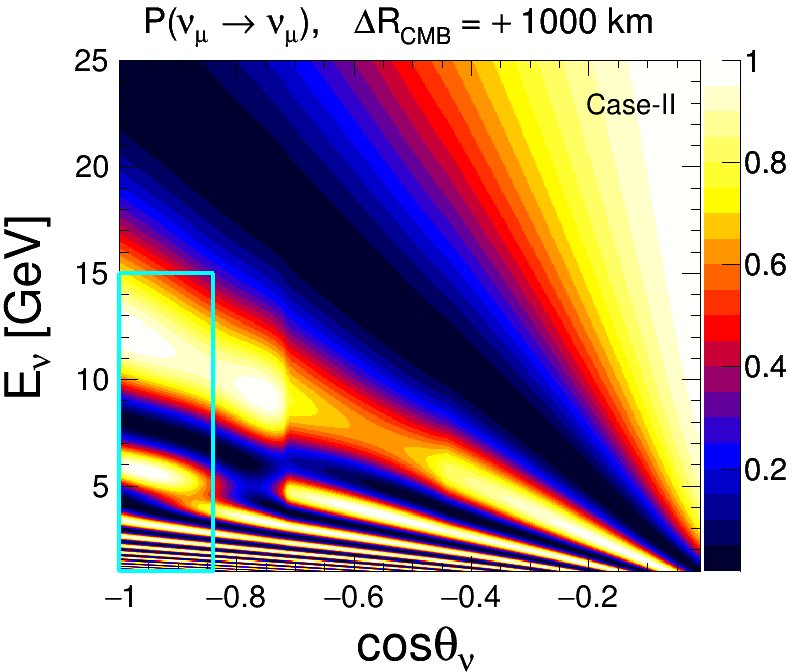}
	\mycaption{Possible impact of larger core radii on three-flavor $\nu_{\mu}\rightarrow\nu_{\mu}$ oscillograms for the three-layered profile of Earth. The oscillograms are shown for $R_\text{CMB} = 3480 \text{ km} + \Delta R_\text{CMB}$ where $\Delta R_\text{CMB} =  (0, +200, +400, +600, +800, +1000)$ km, with the density profile modified according to Case-II. We use three-flavor neutrino oscillation parameters given in Table~\ref{tab:osc-param-value}, where we assume NO and $\sin^2\theta_{23}$ = 0.5.}
	\label{fig:CMB_varying_case_II_lc}
\end{figure}

While showing our results in Sec.~\ref{sec:results}, we have observed in Fig.~\ref{fig:CMB_variation_chisq_margin}  that $\Delta\chi^2_\text{CMB}$ is asymmetric about the standard $R_\text{CMB}$ value in Case-II and Case-III. It also shows non-monotonic behavior at smaller test values of $R_\text{CMB}$. In this appendix, we demonstrate the possible origin of these effects at the probability level in the context of Case-II using Figs.~\ref{fig:CMB_varying_case_II_sc} and \ref{fig:CMB_varying_case_II_lc}, which show the possible impact of smaller and larger core on three-flavor $\nu_{\mu}\rightarrow\nu_{\mu}$ oscillograms. 

In Fig.~\ref{fig:CMB_varying_case_II_sc}, we present $\nu_{\mu}\rightarrow\nu_{\mu}$ oscillograms for smaller core radii with $\Delta R_\text{CMB} =  (-200, -400, -600, -800, -1000)$ km. Note that as $R_\text{CMB}$ decreases (core becomes smaller), the oscillograms show changes in the energy range of $\sim$ 3 to 15 GeV for core-passing neutrinos. We observe that the oscillation patterns for core-passing neutrinos get compressed towards $\cos\theta_\nu = -1$. Moreover, the oscillations in the NOLR/ parametric resonance region becomes more rapid as the core becomes smaller. This oscillatory behavior explains the non-monotonic nature of $\Delta \chi^2_\text{CMB}$ for smaller core.

In Fig.~\ref{fig:CMB_varying_case_II_lc}, we present $\nu_{\mu}\rightarrow\nu_{\mu}$ oscillograms for larger core radii with $\Delta R_\text{CMB} =  (+200, +400, +600, +800, +1000)$ km. Here also, as  as $R_\text{CMB}$ increases (core becomes larger), we see changes in the oscillograms in the energy range of $\sim$ 3 to 15 GeV for core-passing neutrinos. However, here, the oscillation patterns for core-passing neutrinos get stretched towards smaller $|\cos\theta_\nu|$ values. For $\Delta R_\text{CMB} \gtrsim 400$ km, the oscillatory behavior in the NOLR/ parametric resonance region becomes smooth, which accounts for the monotonic behavior of $\Delta \chi^2_\text{CMB}$ at larger core radii.

As a result of the above-mentioned differences between oscillation patterns for smaller and larger core radii, the values of $\Delta \chi^2_\text{CMB}$ are not symmetric about $\Delta R_\text{CMB} = 0$ km, i.e., about the standard $R_\text{CMB}$. The same observations described above for Case-II also hold for Case-III.

\section{Sensitivity Study using the 81-layered PREM Profile}
\label{app:Case-II-prem81}

\begin{figure}
	\centering
	\includegraphics[width=0.49\linewidth]{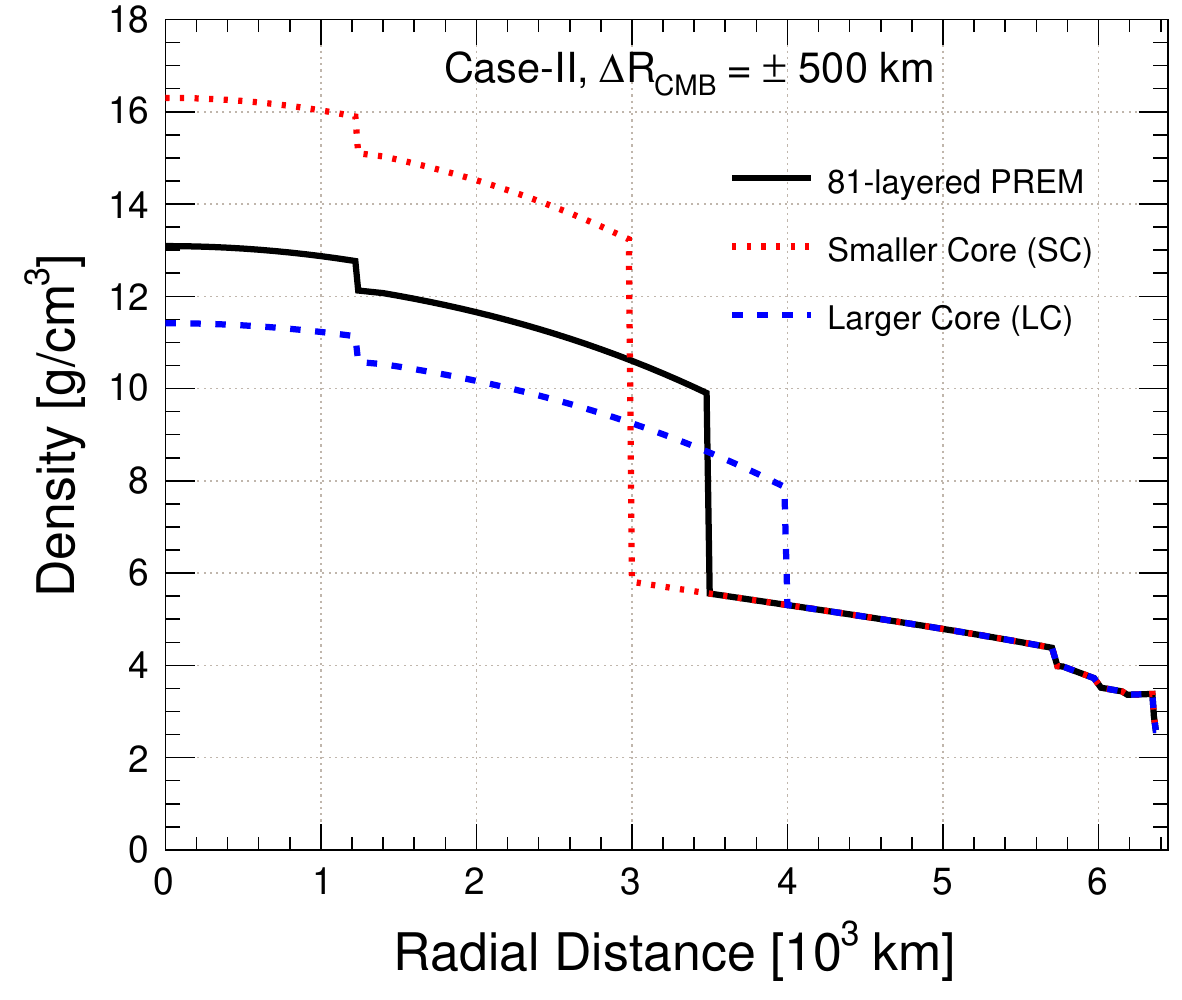}
	\includegraphics[width=0.49\linewidth]{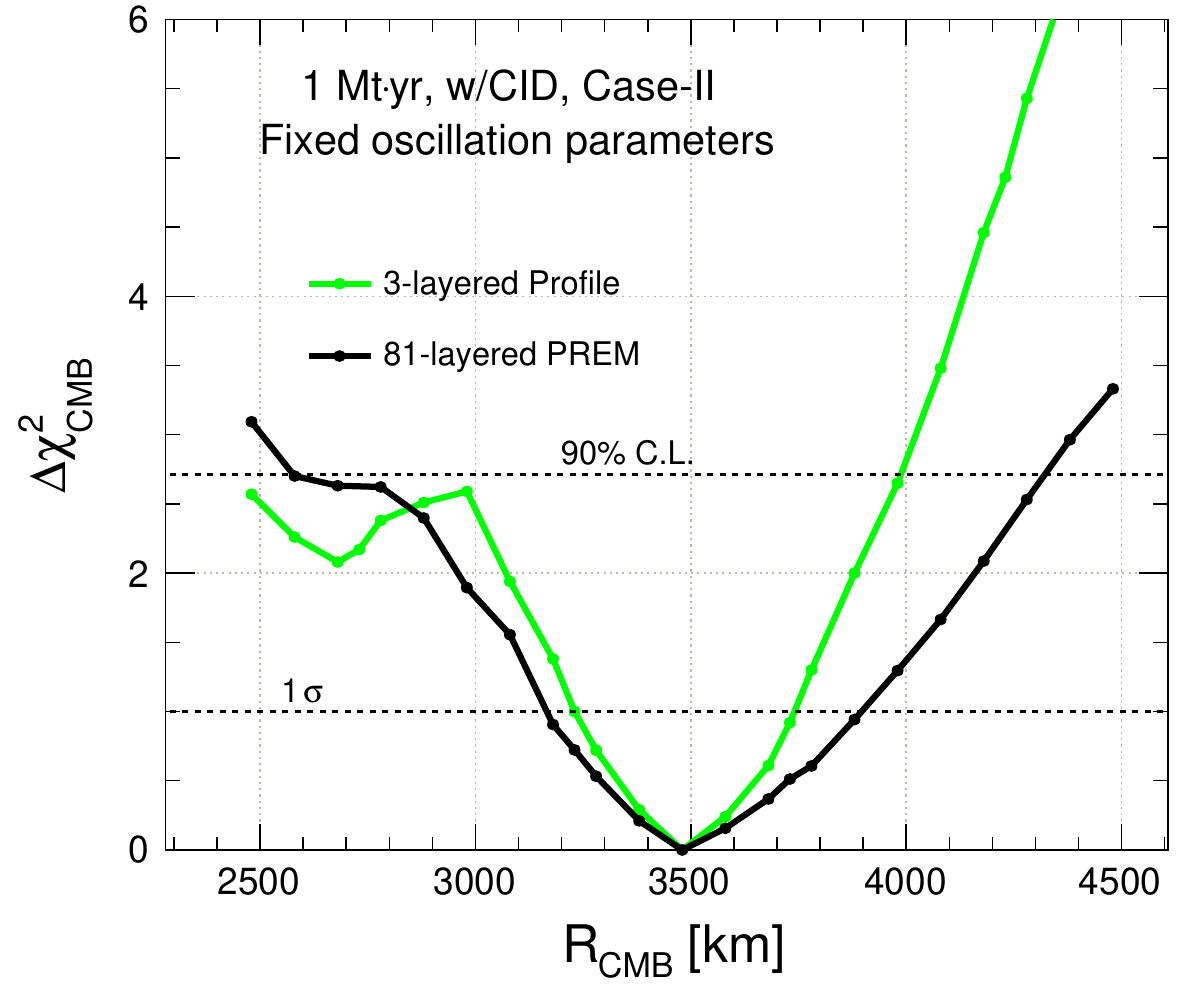} 
	\mycaption{The left panel shows densities as functions of radial distance for modified $R_\text{CMB}$ in the 81-layered PREM profile for Case-II. The black curve shows the standard density profile with $R_\text{CMB}$ = 3480 km. The dotted-red (dashed-blue) curve indicates the density profile in the SC (LC) scenario, where $R_\text{CMB}$ is decreased (increased) by 500 km. The right panel shows the median $\Delta \chi^2_\text{CMB}$ sensitivities as functions of the location of $R_\text{CMB}$ for Case-II. The green and black curves correspond to the sensitivities for the simple three-layered profile and the 81-layered PREM profile, respectively. All the oscillation parameters are kept fixed while evaluating $\Delta \chi^2_\text{CMB}$. The sensitivities correspond to 1 Mt$\cdot$yr exposure where the charge identification capability of ICAL has been used. For simulating the MC data, we use three-flavor neutrino oscillation parameters given in Table~\ref{tab:osc-param-value}, where we assume NO and $\sin^2\theta_{23}$ = 0.5.}
	\label{fig:sensitivity_prem81_case-II}
\end{figure} 

Throughout the paper, we have considered a simple three-layered density profile of Earth to estimate the sensitivity of an atmospheric neutrino experiment like ICAL for locating $R_\text{CMB}$. In this appendix, we explore the effect of going to a more detailed density profile on this sensitivity. We take the Earth density to be according to the 81-layered PREM profile, where the density is interpolated/extrapolated using polynomial functions~\cite{Dziewonski:1981xy}.  For the case of modified $R_\text{CMB}$, the densities of all the layers in the core are scaled by the same fraction, such that the mass of the Earth remains invariant, just like in the Case-II which is discussed in section~\ref{sec:CMB_variation}. The density distribution as a function of radial distance for the 81-layered PREM profile is shown in the left panel of Fig.~\ref{fig:sensitivity_prem81_case-II}, for $\Delta R_\text{CMB} = \pm\,500$ km.  
	
The right panel of Fig.~\ref{fig:sensitivity_prem81_case-II} shows the median sensitivities of the ICAL detector in terms of $\Delta \chi^2_\text{CMB}$ as functions of the location of $R_\text{CMB}$ for Case-II. The value of $R_\text{CMB}$ is modified in small steps of 50 - 100 km up to $\Delta R_\text{CMB} = \pm\,1000$ km with respect to the standard $R_\text{CMB}$ of 3480 km. Since the effects of the uncertainties of oscillation parameters are observed to be negligible in the fit (as can be seen from Table~\ref{tab:results}), we keep all the oscillation parameters fixed in the fit while evaluating $\Delta \chi^2_\text{CMB}$. The figure shows that, with the 81-layered PREM profile, ICAL would be able to measure the location of $R_\text{CMB}$ at $1\sigma$ confidence level with a precision of about $\pm\,350$ km. With the simple three-layered profile, this precision was about $\pm\,250$ km. One of the major reasons for this is the decrease in the density jump at $R_\text{CMB}$ in the 81-layered PREM profile.

\end{appendix}
	
\bibliographystyle{JHEP}
\bibliography{References}

\end{document}